\documentclass[twocolumn]{aastex631}

\usepackage{graphicx}
\usepackage{amsmath}
\usepackage[USenglish]{babel}

\usepackage{tabularx}

\begin{document}

\title{The ALMA Survey of Gas Evolution of PROtoplanetary Disks (AGE-PRO): VIII. \\ The impact of external photoevaporation on disk masses and radii in Upper Scorpius }

\correspondingauthor{Rossella Anania}
\email{rossella.anania@unimi.it}

\author[0009-0004-8091-5055]{Rossella Anania}
\affiliation{Dipartimento di Fisica, Università degli Studi di Milano, Via Celoria 16, I-20133 Milano, Italy}

\author[0000-0003-4853-5736]{Giovanni P. Rosotti}
\affiliation{Dipartimento di Fisica, Università degli Studi di Milano, Via Celoria 16, I-20133 Milano, Italy}

\author[0000-0001-6802-834X]{Mat{\'i}as G{\'a}rate}
\affiliation{Max-Planck-Institut für Astronomie, Königstuhl 17, 69117, Heidelberg, Germany}

\author[0000-0001-8764-1780]{Paola Pinilla}
\affiliation{Mullard Space Science Laboratory, University College London, Holmbury St Mary, Dorking, Surrey RH5 6NT, UK}

\author[0000-0002-4147-3846]{Miguel Vioque}
\affiliation{European Southern Observatory, Karl-Schwarzschild-Str. 2, 85748 Garching bei München, Germany and Joint ALMA Observatory, Alonso de Córdova 3107, Vitacura, Santiago 763-0355, Chile}

\author[0000-0002-8623-9703]{Leon Trapman}
\affiliation{Department of Astronomy, University of Wisconsin-Madison, 475 N Charter St, Madison, WI 53706}

\author[0000-0003-2251-0602]{John Carpenter}
\affiliation{Joint ALMA Observatory, Avenida Alonso de Córdova 3107, Vitacura, Santiago, Chile}

\author[0000-0002-0661-7517]{Ke Zhang}
\affiliation{Department of Astronomy, University of Wisconsin-Madison, 475 N Charter St, Madison, WI 53706}

\author[0000-0001-7962-1683]{Ilaria Pascucci}
\affiliation{Lunar and Planetary Laboratory, The University of Arizona, Tucson, AZ 85721, USA}

\author[0000-0002-2828-1153]{Lucas A. Cieza}
\affiliation{Instituto de Estudios Astrofísicos, Universidad Diego Portales, Av. Ejercito 441, Santiago, Chile}

\author[0000-0002-5991-8073]{Anibal Sierra}
\affiliation{Departamento de Astronom\'ia, Universidad de Chile, Camino El Observatorio 1515, Las Condes, Santiago, Chile}
\affiliation{Mullard Space Science Laboratory, University College London, Holmbury St Mary, Dorking, Surrey RH5 6NT, UK}

\author[0000-0002-2358-4796]{Nicolas T. Kurtovic}
\affiliation{Max-Planck-Institut f\"ur Extraterrestrische Physik, Giessenbachstrasse 1, 85748 Garching, Germany}

\author[0000-0002-1575-680X]{James Miley}
\affiliation{Departamento de Física, Universidad de Santiago de Chile, Av. Victor Jara 3659, Santiago, Chile}
\affiliation{ Millennium Nucleus on Young Exoplanets and their Moons (YEMS), Chile }
\affiliation{Center for Interdisciplinary Research in Astrophysics and Space Exploration (CIRAS), Universidad de Santiago, Chile}

\author[0000-0002-1199-9564]{Laura M. P\'erez}
\affiliation{Departamento de Astronom\'ia, Universidad de Chile, Camino El Observatorio 1515, Las Condes, Santiago, Chile}

\author[0000-0002-1103-3225]{Benoît Tabone}
\affiliation{Institut d'Astrophysique Spatiale, Université Paris-Saclay, CNRS, Bâtiment 121, 91405, Orsay Cedex, France}

\author[0000-0001-5217-537X]{Michiel Hogerheijde}
\affiliation{Leiden Observatory, Leiden University, PO Box 9513, 2300 RA Leiden, the Netherlands}
\affiliation{Anton Pannekoek Institute for Astronomy, University of Amsterdam, the Netherlands}

\author[0000-0003-0777-7392]{Dingshan Deng}
\affiliation{Lunar and Planetary Laboratory, The University of Arizona, Tucson, AZ 85721, USA}

\author[0000-0002-7238-2306]{Carolina Agurto-Gangas}
\affiliation{Departamento de Astronom\'ia, Universidad de Chile, Camino El Observatorio 1515, Las Condes, Santiago, Chile}

\author[0000-0003-3573-8163]{Dary A. Ruiz-Rodriguez}
\affiliation{National Radio Astronomy Observatory; 520 Edgemont Rd., Charlottesville, VA 22903, USA}
\affiliation{Joint ALMA Observatory, Avenida Alonso de Córdova 3107, Vitacura, Santiago, Chile}

\author[0000-0003-4907-189X]{Camilo Gonz\'alez-Ruilova}
\affiliation{Instituto de Estudios Astrofísicos, Universidad Diego Portales, Av. Ejercito 441, Santiago, Chile}
\affiliation{Millennium Nucleus on Young Exoplanets and their Moons (YEMS), Chile and Center for Interdisciplinary Research in Astrophysics}
\affiliation{Space Exploration (CIRAS), Universidad de Santiago, Chile}

\author[0000-0001-9961-8203]{Estephani E. TorresVillanueva}
\affiliation{Department of Astronomy, University of Wisconsin-Madison, 475 N Charter St, Madison, WI 53706}

\begin{abstract}
Protoplanetary disk evolution can be deeply influenced by the UV radiation emitted by neighboring massive stars (mainly of spectral type O and B). We show that the process of \textit{external photoevaporation}, which causes an outside-in depletion of disk material due to environmental UV radiation, can lead to a significant decrease in disk size, and moderate in disk mass and lifetime even at moderate irradiation levels (1-10 G$_{0}$). In this work we investigate the role of external photoevaporation in shaping the masses and sizes of the ten AGE-PRO disks in the Upper Scorpius region, which we estimate to be subject to FUV fluxes ranging between $\sim$2 and $\sim$12 G$_{0}$, on average. 
We compare the disk masses and sizes resulting from 1D numerical viscous evolution simulations in which the effect of external photoevaporation is included, to the values retrieved from the AGE-PRO observations.
While the pure viscous framework fails in adequately explaining the observed disk properties in Upper Scorpius, with the inclusion of external photoevaporation we can successfully reproduce gas disk sizes for 7 out of 10 sources within a factor $<$2, when the initial disk mass is 1-10\% of the stellar mass.
We emphasize the importance of accounting for the environmental irradiation when comparing star-forming regions of different ages, even when moderate FUV irradiation fields are experienced, as in the case of Upper Scorpius.
\end{abstract}

\keywords{} 

\section{Introduction} \label{sec:intro}


Protoplanetary disks form around newly born stars and evolve in response to various physical processes.
Many of these processes are \textit{internal} to the stellar-disk system. Among them, there are the classical viscous framework, driven by magneto-rotational instabilities (MRI), 
(\citeauthor{LeP_solution} \citeyear{LeP_solution}, \citeauthor{pringle} \citeyear{pringle}, \citeauthor{Balbus_mri} \citeyear{Balbus_mri}), and the
accretion driven by the depletion of material through magneto-hydrodynamical (MHD) winds (e.g. \citeauthor{Bai_Stone_2013_MHD} \citeyear{Bai_Stone_2013_MHD},  \citeauthor{Tabone_mhd} \citeyear{Tabone_mhd}).
Furthermore, thermal photoevaporative winds, driven by X-ray and Ultraviolet (UV) radiation from the central star, contribute to the loss of material (\citeauthor{Clarke_2001}  \citeyear{Clarke_2001}, \citeauthor{Ercolano_2008} \citeyear{Ercolano_2008}, \citeauthor{Picogna_2019} \citeyear{Picogna_2019}, \citeauthor{Sellek_2022} \citeyear{Sellek_2022}).
In the viscous framework the observed disk size is expected to increase with time, while in the wind-driven scenario, depending on the amount of material lost in winds, the disk size can either remain constant or decrease (\citeauthor{Trapman_mhd} \citeyear{Trapman_mhd}). Consequently, measuring disk sizes is a powerful tool for constraining the dominant process driving disk evolution (and accretion). 
The combined efforts in measuring disk sizes, investigating disk winds\citep{Pascucci_2023}, and the comparison with models, are crucial to give insights into the primary driver of accretion.
\\

In addition to the processes mentioned above, two \textit{external} processes play a crucial role in shaping disk properties: star-disk interactions and external photoevaporation.
Gravitational interactions with other stars (in bound or unbound systems) may perturb and/or truncate the disk, steepening the surface density profile and altering disk properties (e.g. \citeauthor{Rosotti_2014_stardiscinteraction} \citeyear{Rosotti_2014_stardiscinteraction}, \citeauthor{Breslau_2014_stardiscinteraction} \citeyear{Breslau_2014_stardiscinteraction}, \citeauthor{Cuello_2023} \citeyear{Cuello_2023}). However, the majority of young stars is expected to form in dense and massive clusters or associations \citep{Fatuzzo_Adams_2008}, where external photoevaporation is the dominant environmental factor
acting on protoplanetary disks (e.g. \citeauthor{Winter_2018} \citeyear{Winter_2018}, \citeauthor{Concha-Ramirez_2023} \citeyear{Concha-Ramirez_2023}), induced by the intense UV radiation emitted by massive stars (mostly of spectral type O and B).
The FUV (far-ultraviolet) and EUV (extreme-ultraviolet) radiation heats the outer, less gravitationally bound, disk region causing an outside-in depletion of material (\citeauthor{Johnstone_1998} \citeyear{Johnstone_1998}, \citeauthor{Adams_2004} \citeyear{Adams_2004}).
This process affects disk mass through material loss in thermal winds and, more significantly, influences disk size and lifetime.
Indeed, since the surface density profile is truncated and the disk cannot spread as in the viscous case, the outer disk boundary is modified and disk dispersal is accelerated (\citeauthor{Clarke_2007} \citeyear{Clarke_2007}, \citeauthor{Anderson_adams_2013} \citeyear{Anderson_adams_2013}, \citeauthor{Coleman_2022} \citeyear{Coleman_2022}).
Moreover, the process of planet formation can also be affected, since the disk truncation limits the pebble mass reservoir for planet growth \citep{Quiao_coleman_2023}. 

Observations in dense and highly irradiated clusters (e.g. Orion Nebula Cluster, Carina), unveil direct evidence of externally photoevaporating protoplanetary disks. These disks, known as \textit{proplyds}, are surrounded by bright ionization fronts that are oriented towards the nearest massive star
(e.g. \citeauthor{ODell_1993} \citeyear{ODell_1993}, \citeauthor{Smith_carina_proplyds} \citeyear{Smith_carina_proplyds}).
Moreover, indirect evidence demonstrates the influence of external photoevaporation on disk properties, such as the inverse correlation between the number of massive stars and the disk fraction observed in Cygnus \citep{Guarcello_cygnus}, and the disk dust mass and FUV flux in the L1641 and L1647 regions of Orion \citep{SODA_2023} as well as in the $\sigma$-Orionis cluster (\citeauthor{Ansdell_sigma_orionis} \citeyear{Ansdell_sigma_orionis}, \citeauthor{Mauco_sigmaori} \citeyear{Mauco_sigmaori}) and in the ONC (\citeauthor{Eisner_2018} \citeyear{Eisner_2018}, \citeauthor{Otter_2021} \citeyear{Otter_2021}). 

However, the regions described above are characterized by an intense FUV irradiation field ($> 10^{3} \ \rm{G}_{0}$). The role of a moderate FUV flux (1-100 G$_{0}$) on disk properties, which is a more common scenario in the most nearby regions (at a distance $<$ 200 pc from the Sun), is still to be quantified.
For this purpose Upper Scorpius (Upper Sco) is a perfect laboratory.

Upper Sco, located at $\sim$142 pc from the Sun \citep{Fang_pascucci_2023}, is the oldest nearby Star-Forming Region (SFR), aged $\sim$3-14 Myr (e.g. \citeauthor{ScoCen_sub_clusters} \citeyear{ScoCen_sub_clusters})\footnote{In this work we refer to Upper Sco as a star-forming region for the presence of young stars even if, due to its age, the process of star formation has presumably stopped}. It exemplifies the late stages of (Class II) disk evolution, with disks that have survived both accretion and environmental processes.
Upper Sco includes $\sim$300 stars presenting full, evolved, transition or debris disks (Carpenter et al. 2024, sub.). 
This region is a sub-group of the Scorpius-Centaurus (Sco-Cen) OB association, which 
counts $\sim$150 OB stars, $\sim$50 of which are in Upper Sco \citep{Luhman_uppersco_census}. Therefore, Upper Sco disks may have been strongly influenced by the prolonged exposure to external UV across their evolution. As a consequence, a sample of disks in the Upper Sco region may not be representative of the evolved phase of disks in low-irradiated SFRs, such as Taurus or Lupus, and should be modeled separately \citep{Zagaria_2023_UpperSco}.

The study from \citet{Trapman_2020} showed preliminary evidence that the gas disks in the Upper Sco region appear to be smaller than predicted by viscous models. Moreover, the work of \citet{Trapman_mhd} indicates that even MHD wind-driven models fail to fully explain gas disk sizes in Upper Sco: for this scenario to reproduce the observed disk sizes, disks at younger ages would need to have larger gas-to-dust size ratios than the values typically observed in young regions.

A potential solution to this problem might involve accounting for the external photoevaporation process when comparing models to observations. \\

The ALMA survey of Gas Evolution in PROtoplanetary disks (AGE-PRO; ID 2021.1.00128.L) is designed to trace the evolution of gas disk mass and size throughout the lifetime of protoplanetary disks. In order to investigate the three main stages of disk life (embedded disk phase, middle age, and late stage), the disk sample is selected from three nearby SFRs of different ages: Ophiuchus \citep{AGEPRO_II_Ophiuchus}, Lupus \citep{AGEPRO_III_Lupus}, and Upper Sco \citep{AGEPRO_IV_UpperSco}.
In this study we focus on the ten AGE-PRO targets in Upper Sco (see \citeauthor{AGEPRO_I_overview} \citeyear{AGEPRO_I_overview} and \citeauthor{AGEPRO_IV_UpperSco} \citeyear{AGEPRO_IV_UpperSco} for additional details on the sample selection).
\\

The aim of this work is to test if 1D viscous evolution models with the inclusion of external photoevaporation are able to explain disk masses 
and radii observed by the AGE-PRO survey in the Upper Sco region. 
This study allows us to investigate the role of external photoevaporation on disk properties when disks are subject to moderate and low levels of irradiation (1-10 G$_{0}$).

The sample of disks analysed in this work is detailed in Section \ref{sec: disk_sample}.
In Section \ref{sec: FUV calculation} we computed the average FUV flux experienced by the ten AGE-PRO targets in UpperSco. In Section \ref{sec:disc_evolution_model} we describe the analytic disk model and the simulation setup adopted. In Section \ref{sec:results} we show the outcomes of our simulations, and compare the disk masses and radii with the AGE-PRO observations. Our results, and additional processes affecting disk masses and sizes, are discussed in Section \ref{sec:discussion}, while Section \ref{sec:conclusion} contains our final remarks.
\section{Disk sample}\label{sec: disk_sample}
\begin{table*}[ht]
    \centering
    \caption{Upper Sco stellar and disk parameters}
    \def\arraystretch{1.2}
    \begin{tabularx}{1.9\columnwidth}{ccccccccc}
    \hline
    \hline
    2MASS id & AGE-PRO id 
    & dist & M$_{\mathrm{\star}}$ & age & log M$_{\mathrm{gas}}$ & F$_{\mathrm{mm}}$ & R$_{^{12}\mathrm{CO}, 90\%}$ & R$_{\mathrm{dust}}$ \\
    & & [pc] & [$\mathrm{M_{\odot}}$] & Myr & $[\mathrm{M}_{\odot}]$ & [mJy] & [AU] & [AU] \\
    \hline
    J16120668-3010270 & UppSco 1  
    & 131.9
    & 0.51 & 4.37$^{+ 4.75}_{- 2.23}$ & -2.49$^{+ 0.15}_{- 0.11}$ & 34.27$\pm$0.05 & 166.9$^{+5}_{-1.8}$ & 87.64 $\pm$0.62 \\ 
    J16054540-2023088 & UppSco 2  & 137.6 &
    0.13 & 2.09$^{+ 3.80}_{- 1.30}$ & -3.9$^{+0.38}_{-0.24}$ & 7.03$\pm$0.03 & 50.6$^{+10.7}_{-23.4}$ & $<$ 17\\ 
    J16020757-2257467 & UppSco 3  & 139.6 &
    0.37 & 3.72$^{+ 5.19}_{- 2.13}$ & -3.19$^{+0.24}_{-0.43}$ & 2.99$\pm$0.04 & 34.1$^{+3.1}_{-1.9}$ & 27.0$\pm$6.6\\
    J16111742-1918285 & UppSco 4  & 136.9 &
    0.50 & 2.09$^{+ 1.80}_{- 1.01}$ & -3.69$^{+0.92}_{-1.33}$ & 0.208$\pm$0.02 & 46.3$^{+1.1}_{-1.4}$ & $<$ 56 \\ 
    J16145026-2332397 & UppSco 5  & 143.9 &
    0.29 & 3.8$^{+ 6.91}_{- 2.38}$ & -5.36$^{+0.27}_{-0.2}$ & 0.68$\pm$0.03 & 30.8$^{+1.7}_{-3.4}$ & $<$ 19\\
    J16163345-2521505 & UppSco 6  & 158.4 &
    0.52 & 5.89$^{+ 7.29}_{- 3.19}$ & -4.16$^{+0.33}_{-0.33}$ & 1.91$\pm$0.03 & 159.7$^{+ 15.2}_{-10.3}$ & 63.0$\pm$19.0  \\ 
    J16202863-2442087 & UppSco 7  & 152.7 & 
    0.34 & 1.78$^{+ 1.69}_{- 0.91}$ & -3.38$^{+0.29}_{-0.23}$ & 3.14$\pm$0.03 & 160.5$^{+1.2}_{-1.7}$ & 47.0$\pm$22.0 \\ 
    J16221532-2511349 & UppSco 8  & 138.9 &
    0.30 & 2.4$^{+ 3.09}_{- 1.44}$ &  -2.41$^{+0.27}_{-0.23}$ & 22.21$\pm$0.05 & 144.1$^{+1.5}_{-2.8}$ & 27.19$\pm$0.18 \\
    J16082324-1930009 & UppSco 9  & 136.9 &
    0.56 & 4.68$^{+ 4.87}_{- 2.33}$ & -1.28$^{+0.2}_{-0.38}$ & 34.86$\pm$0.12 & 180.8$^{+4.6}_{-8.7}$ & 47.8$\pm$0.34 \\ 
    J16090075-1908526 & UppSco 10 & 136.9 & 
    0.53 & 2.34$^{+ 1.92}_{- 1.05}$ &  -2.14$^{+0.43}_{-0.45}$ & 38.04$\pm$0.13 & 74.6$^{+ 1.8}_{-3.7}$ & 42$\pm$1.5 \\
    \hline
    \end{tabularx}
    \begin{minipage}{1.8\columnwidth}
    \vspace{0.1cm}{\footnotesize{\textbf{Notes:} For each of the AGE-PRO targets in Upper Sco, the table indicates (from left to right) 2MASS identifier, AGE-PRO identification name, distance \citep{Bailer-Jones},  stellar mass, isocronal age (\citeauthor{AGEPRO_I_overview}\citeyear{AGEPRO_I_overview}), gas disk mass \citep{AGEPRO_V_gas_masses}, mm flux \citep{AGEPRO_IV_UpperSco}, gas disk radius \citep{AGEPRO_XI_gas_disk_sizes}, and dust disk radius \citep{AGEPRO_X_dust_disks} for which UppSco 2, UppSco 3, and UppSco 5 are upper limits.} }
    \end{minipage}
    \label{tab:source_params}
\end{table*}
The source sample investigated in this study consists of the ten Upper Sco protoplaneraty disks targeted by the AGE-PRO ALMA Large Program. Table \ref{tab:source_params} summarises stellar parameters and dust and gas disk properties as retrieved from AGE-PRO.
The targets are selected to have previous detection of mm continuum and $^{12}$CO line emission, and stellar masses ranging from $\sim$0.1 M$_{\odot}$ and $\sim$0.6 M$_{\odot}$ \citep{AGEPRO_I_overview}. 
For each source, the stellar mass is obtained from stellar evolution models (\citeauthor{Feiden_masses} \citeyear{Feiden_masses}, \citeauthor{baraffe} \citeyear{baraffe}), and the corresponding isocronal age is also computed. 
A more detailed description of the source sample, along with the method used for measuring disk masses and sizes can be found in 
\citet{AGEPRO_IV_UpperSco}, \citet{AGEPRO_V_gas_masses}, \citet{AGEPRO_XI_gas_disk_sizes}, and \citet{AGEPRO_X_dust_disks}.
In this work we compare the results of numerical simulations with the disk properties listed in Table \ref{tab:source_params}.
\\

\section{Calculation of the FUV flux} \label{sec: FUV calculation}
In this Section we provide the technique used to estimate the external FUV flux at the position of the ten AGE-PRO sources in Upper Sco. 
Since the majority of the massive stars in Upper Sco are late-type B, with no early-type O star in the proximity, we expect the FUV flux to dominate mass loss over the disk lifetime against EUV flux \citep{winter_photoev}. Therefore, we focus only on this irradiation regime, where the range of wavelengths corresponding to far-ultraviolet is [912 - 2400] $\mathring{\rm{A}}$.
First we select the O and B stars in the neighborhood of each source (Sec. \ref{subsec:field_star_selection}). Then, we calculate the FUV flux as the contribution of all the selected field stars, taking into account the uncertainties introduced when working with 3D distances (Sec. \ref{subsec:FUV_calculation}). 
\subsection{Field Stars Selection}\label{subsec:field_star_selection}
    \begin{figure*}[htbp]
    \centering
    
    \begin{minipage}[b]{0.45\textwidth}
        \centering
        \includegraphics[width=\textwidth]{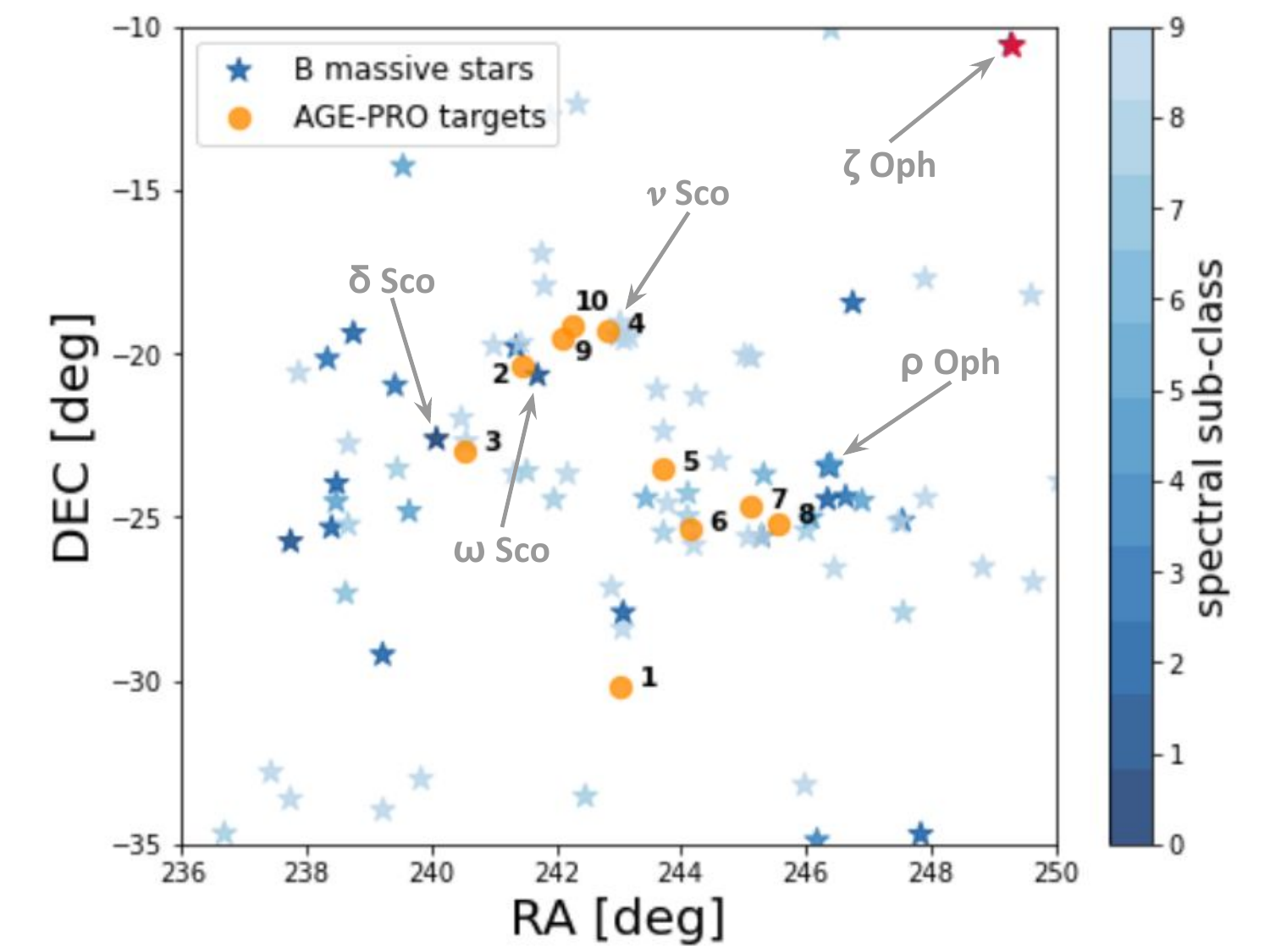}
    \end{minipage}
    \hfill
    \hspace{0.002\textwidth}
    \begin{minipage}[b]{0.50\textwidth}
        \centering
        \includegraphics[width=\textwidth]{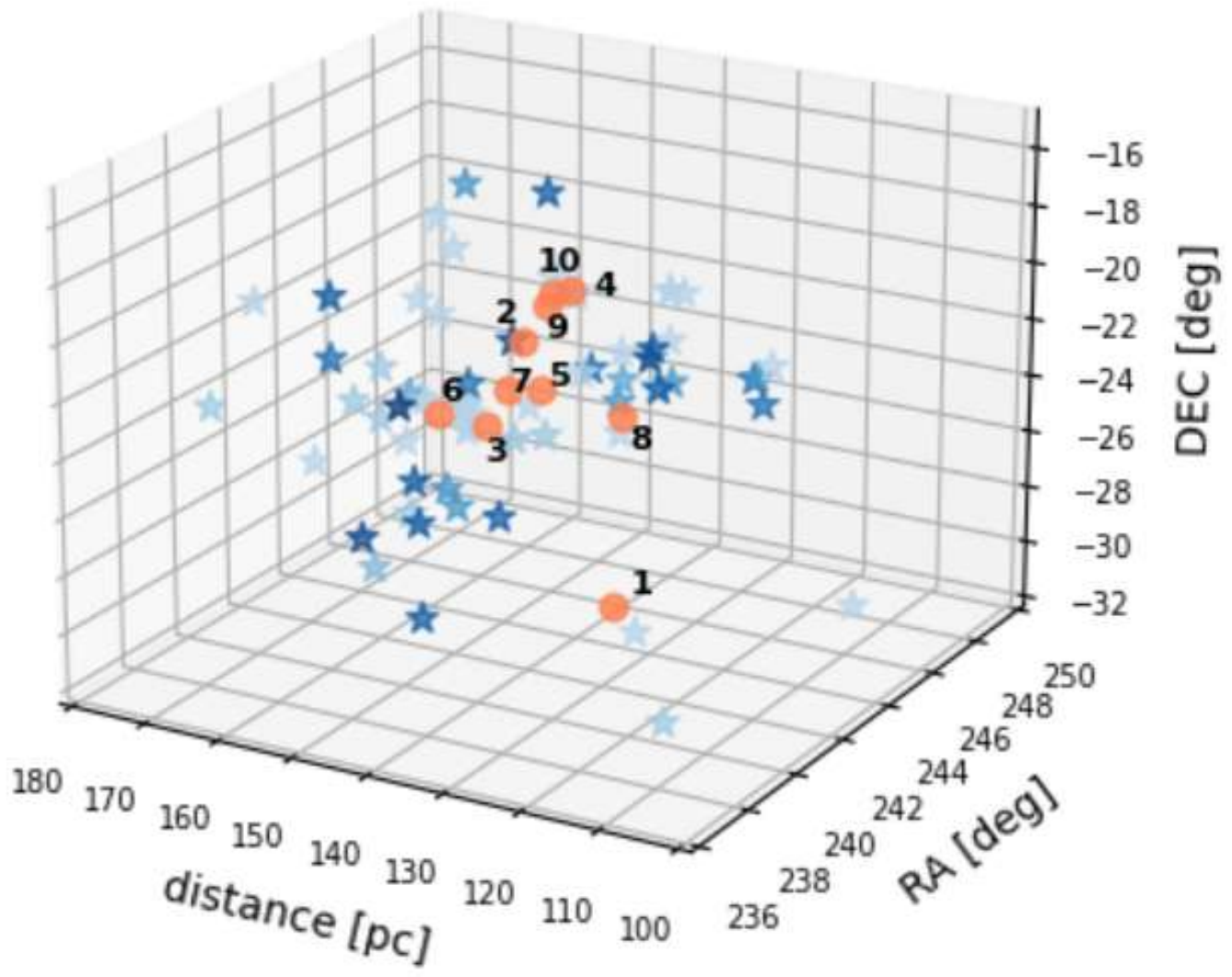}
    \end{minipage}
    \caption{Distribution of the ten AGE-PRO sources in Upper Sco and the O and B stars closest to them. B-type stars are color-coded based on their spectral sub-class (from B0 to B9). The only O-type star survived in the considered region is \emph{$\zeta$ Oph}, flagged in red. \emph{Left}: projected-sky map.  \emph{Right}: 3D map.}
    \label{fig:skymap}
    \end{figure*}
    From \citet{Luhman_uppersco_census} and \citet{Luhman_uppsco} we can retrieve a list of OB stars members of the Upper Sco region (49 stars). To avoid excluding any relevant stars from the calculation, we enlarge the sample to cover all OB stars in a $\sim$55 degree-wide box that is centred on our source sample. 
    This box size is large enough to capture meaningful FUV flux contributions (i.e. $>$ 0.1 G$_{0}$) from nearby and distant OB stars and contains in total 174 objects (see Appendix \ref{appendix:ob_stars} for the complete list of OB stars).    
    Early A-type stars do not contribute significantly to the total FUV flux \citep{cleeves_imlup} and therefore are not included in our calculation.
    The stellar parameters are taken from Gaia DR3 (\citeauthor{Gaia_dr3} \citeyear{Gaia_dr3}, \citeauthor{Gaia_dr3_1} \citeyear{Gaia_dr3_1})\footnote{\url{https://www.cosmos.esa.int/web/gaia/dr3}} and Hipparcos catalog \citep{Hipparcos_catalogue}\footnote{\url{ https://www.cosmos.esa.int/web/hipparcos/home} } when Gaia data are not available for the brightest sources. Afterwards, we searched in the ALS II catalog \citep{ALSII_catalogue} for any star included the previously defined box, but missing in the previous selection (due to unclear or undefined spectral type).
    We retrieve the spectral type and luminosity class of the stars from the SIMBAD database \citep{Simbad_catalogue}. Moreover, we added to our selection the hot sources according to the Gaia Blue-Red pass band spectra (those with reported effective temperature in Gaia but missing spectral type in SIMBAD).
    The extended list of selected massive stars, along with their parameter, is discussed in Appendix \ref{appendix:ob_stars}.
    
    Figure \ref{fig:skymap} illustrates the distribution of the ten AGE-PRO sources in Upper Sco and the OB stars closest to them, using a projected sky map (left panel) and a 3D sky map (right panel). The field is dominated by B9/B8 stars. \emph{$\zeta$ Oph}, the only O-type star (in red in Figure \ref{fig:skymap}), contributes with less than $0.5 \ \mathrm{G}_{0}$ to the total FUV flux on the AGE-PRO sources.
    The distance of the considered objects is determined through parallax inversion (we motivate this approach in Appendix \ref{Appendix:flux_error}).

    \subsection{FUV Flux Experienced by the Sources}\label{subsec:FUV_calculation}
    The FUV flux experienced by each disk is evaluated by adding the contribution of all the selected field stars.
    Spectral types and luminosity classes of the OB stars are derived from the SIMBAD database and subsequently linked to effective temperatures, thereby distinguishing between Main-Sequence stars, Giant and Super-Giants using the tables from \citet{Teff} and \citet{teff_giants}. For simplicity we are assuming a blackbody emission for the stars. While this tends to underestimate the total stellar flux, we expect the uncertainty to be dominated by the parallax measurements (see Appendix \ref{Appendix:flux_error}).
    Given the effective temperature, we fit the stellar masses using the stellar evolution models of \cite{stellar_evol_models_B}. Subsequently, we derive the FUV luminosity from the stellar mass using the empirical relationship derived by \cite{winter_photoev}.
    Finally, the FUV flux experienced by each disk is determined by the contribution of the overall field stars: 
    \begin{equation}
        F_{\mathrm{FUV}} = \sum_{\mathrm{B}} \frac{L_{\mathrm{FUV},B}}{4\pi |x_{\mathrm{disk}} - x_{\mathrm{B}}|^{2}},
        \label{eq:fuv_flux}
    \end{equation}
    where $L_{\mathrm{FUV,B}}$ is the FUV luminosity of the massive star and $|x_{\mathrm{disk}} - x_{\mathrm{B}}|$ is the relative distance of the disk from the massive star. It is worth noticing that in our approach we are neglecting the influence of the extinction from residual gas and dust on shielding protoplanetary disks from the FUV irradiation.
    On the other hand, recent studies have shown that in massive star forming regions where O-type stars are present, this effect can be relevant only within the first half million year of stellar evolution (\citeauthor{Ali_2019} \citeyear{Ali_2019}, \citeauthor{Wilhelm_shielding} \citeyear{Wilhelm_shielding}). Since the AGE-PRO sample in Upper Sco has age spread ranging from $\sim$2 to $\sim$6 Myr (Table \ref{tab:source_params}), we do not expect this effect to have a significant impact on our flux calculation. In Section \ref{sec:discussion} we analyze the case in which the external irradiation acts on disks after a certain shielding time.
    
    In order to account for and quantify the error on the flux value arising from uncertainties in parallax measurements of the stellar objects, we use a Monte Carlo method. Specifically, the parallaxes of the single objects are randomly sampled assuming a normal distribution around the Gaia measurements, with standard deviation given by Gaia parallax uncertainties, and recompute the flux using the new values. We repeat this operation enough times to extract the best flux value as the median of the distribution and the associated uncertainty (i.e. we stopped the iteration once the median flux stabilized, even with an increased number of iterations). A further discussion about the results obtained using this approach, as well as an estimate of the upper limit of the FUV irradiation field experienced by each source, is given in Appendix \ref{Appendix:flux_error}.
    Figure \ref{fig:fuv_flux} shows for each source the median of the flux value and the relative uncertainties. The error bars reach down to the 16$^{\mathrm{th}}$ percentile and up to the 84$^{\mathrm{th}}$ percentile of the flux distribution. The extended distribution of FUV fluxes is shown in Appendix \ref{appendix:flux_distr}, while the median FUV flux (input parameter in our simulations) and the uncertainties are listed in Table \ref{tab:mstar_fuv}.    
    In this calculation we are employing the current 3D separation between disks and massive stars (obtained by inverting the parallax from Gaia measurements), which is not a static parameter in time and might have influenced the experienced FUV field. We discuss this last topic in Section \ref{subsec: discussion_additional_factors}.

    \begin{figure}[htbp]
    \centering
    \includegraphics[width=0.45\textwidth]{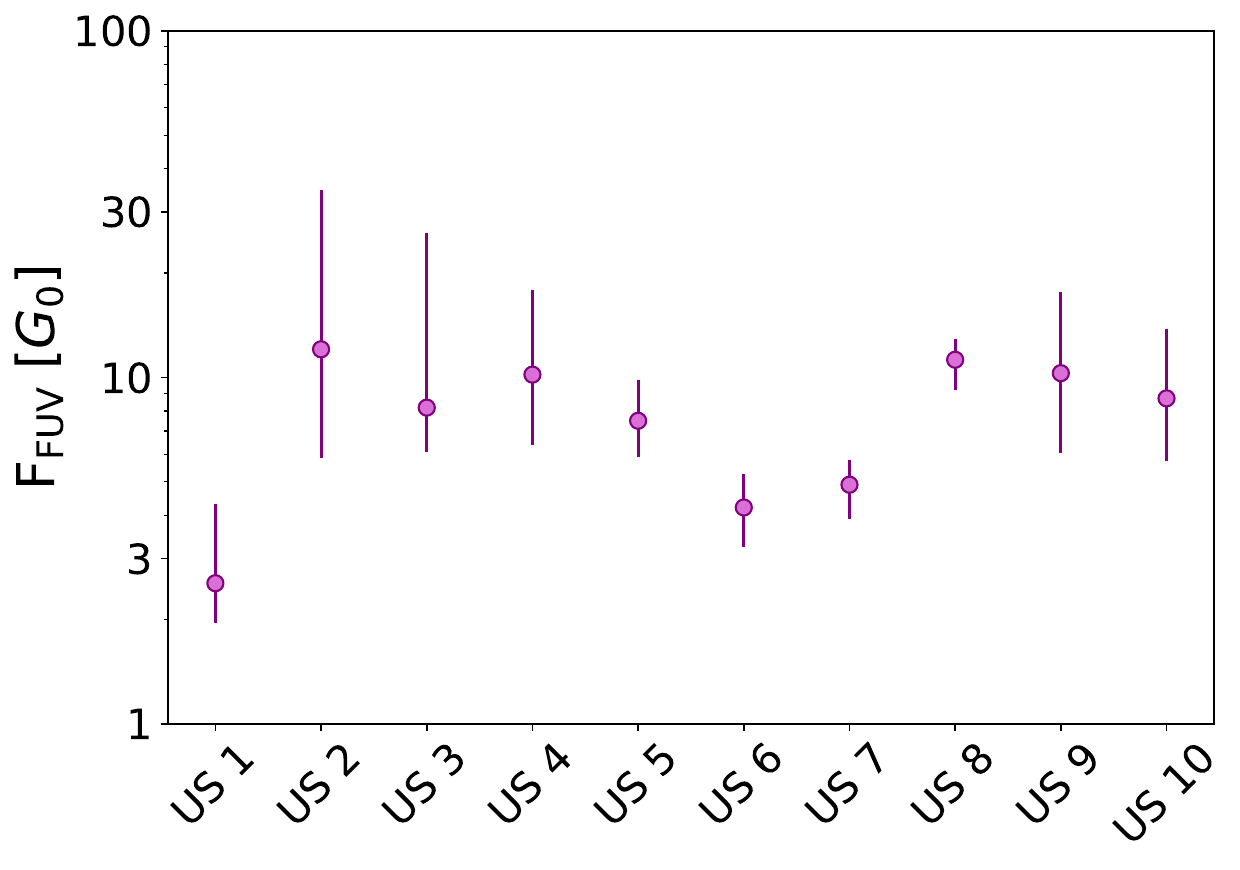}
    \caption{FUV fluxes experienced by the ten AGE-PRO targets in Upper Sco. Error bars extend from the 16$^{\mathrm{th}}$ percentile to the 84$^{\mathrm{th}}$ percentile of the distribution, while purple dots refers to the median values.}
    \label{fig:fuv_flux}
    \end{figure}
\begin{table}[htb]
    \centering
    \caption{Upper Sco source parameters and FUV flux}
    \def\arraystretch{1.2}
    \begin{tabular*}{0.8\columnwidth}{ccccccc}
    \hline
    \hline
    AGE-PRO id 
    & $F_{\mathrm{FUV}}$ & $F_{\mathrm{FUV, 16 \%}}$ & $F_{\mathrm{FUV, 84 \%}}$ \\
    & [$\mathrm{G_{0}}$] & [$\mathrm{G_{0}}$] & [$\mathrm{G_{0}}$] \\
    \hline

    UppSco 1  & 2.546 & 1.979 & 4.256 \\ 
    UppSco 2  & 12.049 & 5.898 & 34.335 \\ 
    UppSco 3  & 8.176 & 6.172 & 25.786 \\
    UppSco 4  & 9.182 & 6.434 & 17.632 \\ 
    UppSco 5  & 7.493 & 5.934 & 9.741 \\
    UppSco 6  & 4.212 & 3.262 & 5.202 \\ 
    UppSco 7  & 4.898 & 3.952 & 5.721 \\ 
    UppSco 8  & 10.241 & 9.281 & 12.771 \\
    UppSco 9  & 9.781 & 6.095 & 17.423 \\ 
    UppSco 10 & 8.696 & 5.786 & 13.691 \\
    \hline
    \end{tabular*}
    \begin{minipage}{0.7\columnwidth}
    \vspace{0.1cm}{\footnotesize{\textbf{Notes:} For each AGE-PRO target listed in Column 1, we indicate the median computed FUV flux and uncertainties corresponding to the 16$^{\mathrm{th}}$ and 84$^{\mathrm{th}}$ percentiles.} }
    \end{minipage}
    \label{tab:mstar_fuv}
\end{table}

\section{disk evolution model}\label{sec:disc_evolution_model}

\subsection{Gas Evolution}
In this work we consider a geometrically thin and axisymmetric disk undergoing viscous accretion and photoevaporative mass loss in its outer region. The gas surface density $\Sigma_{\mathrm{g}}(R,t)$ is obtained by solving the 1D standard diffusion equation (\citeauthor{LeP_solution},\citeyear{LeP_solution}; \citeauthor{pringle},\citeyear{pringle}):
\begin{equation}
    \frac{\partial}{\partial t} \Sigma_{\mathrm{g}} = \frac{3}{R} \frac{\partial}{\partial R} \left( R^{1/2} \frac{\partial}{\partial R} (\nu \Sigma_{\mathrm{g}} R^{1/2})  \right) - \dot{\Sigma}_{\mathrm{ext}},
    \label{eq:diffusion_eq}
\end{equation}
where $\nu = \alpha c_{\rm{s}} H_{\rm{g}}$ is the kinematic viscosity, depending on the gas scale height $H_{\rm{g}}$ and the sound speed $c_{\rm{s}}$, while $\dot{\Sigma}_{\mathrm{ext}}$ accounts for the rate of mass lost in external photoevaporative winds. We assume the constant $\alpha$ viscosity prescription of \citet{alpha_prescription}. \\

The radial temperature profile is set considering the approximation for a flared disk, passively heated by the central star under an angle $\psi$:
\begin{equation}
    T(R) = \left( \frac{\psi L_{\star}}{8 \pi R^{2} \sigma_{\rm{SB}}}\right)^{1/4},
    \label{eq:temp_profile}
\end{equation}
where $\sigma_{\rm{SB}}$ is the Stefan-Boltzmann constant, and $L_{\star}$ is the stellar luminosity. We use a constant flaring irradiation angle, $\psi = 0.05$, leading to the temperature scaling with radius as $T(R) \propto R^{-1/2} $.\\

In our model we neglect the effects of internal photoevaporation, since the goal of this work is to focus on the impact of a moderate external irradiation. Furthermore, external photoevaporation impacts disk evolution across the entire disk life, while  internal photoevaporation is particularly relevant for disk dispersal at the end of disk life, on a timescale that is a small fraction of the disk lifetime (e.g. \citeauthor{Clarke_2001} \citeyear{Clarke_2001}). \\  

In order to estimate the mass loss rate arising from external photoevaporation, 
we perform a bi-linear interpolation of a sub-set of the FRIEDv2 grid \citep{FRIEDv2} (i.e. the updated version of the FRIED grid for mass loss rate \citeauthor{fried}\citeyear{fried}), considering a fiducial ratio between the Polycyclic Aromatic Hydrocarbon (PAH) abundance and the dust abundance of 1, corresponding to ISM-like PAHs for ISM-like dust, and the \textit{grain growth} model as grain growth is expected to happened in Upper Sco and wind is dust-depleted 
(\citeauthor{facchini_2016},\citeyear{facchini_2016}; \citeauthor{fried},\citeyear{fried}). 
The bilinear interpolation is performed using, for a given stellar mass and FUV flux, the disk surface density at each radial position
to interpolate through the grid. We identify as \emph{truncation radius}, $R_{\mathrm{t}}$, the radius associated with the maximum mass loss rate, which corresponds to the position of the optically thin/thick transition of the wind at the FUV wavelengths. Results from numerical simulations support the fact that the photoevaporation rate increases in the outer, weakly bound, disk regions if the wind is optically thick in the FUV, and decrease with radius in the optically thin region (\citeauthor{sellek_fried} \citeyear{sellek_fried}, \citeauthor{FRIEDv2} \citeyear{FRIEDv2}).
Following the numerical approach of \citet{sellek_fried}, we implement the outside-in depletion of gas and dust at $R > R_{\mathrm{t}}$ weighting on the total mass external to the truncation radius, $M(R>R_{\mathrm{t}})$, such that the mass loss at a grid cell $i$ is:
\begin{equation}
    \dot{M}_{\mathrm{ext},i} = \dot{M}_{\mathrm{tot}} \frac{M_{i}}{M(R>R_{\mathrm{t}})},
\end{equation}
where $\dot{M}_{\mathrm{tot}}$ is the total mass loss rate outside the truncation radius.

\subsection{Dust Evolution}
The dust evolution is modeled considering a distribution of particle species, with different mass and size, undergoing the processes of dust transport, dust growth and dust entrainment with photoevaporative wind.
  
Referring to a particular dust species, the dust transport is described by the advection-diffusion equation \citep{advection_diffusion}:
\begin{multline}
     \frac{\partial}{\partial t}(R \Sigma_{\mathrm{d}}) + \frac{\partial}{\partial R}(R \Sigma_{\mathrm{d}}\nu_{\mathrm{d}}) - \frac{\partial}{\partial R} \left( R D_{\mathrm{d}} \Sigma_{\mathrm{g}} \frac{\partial}{\partial R} \left( \frac{\Sigma_{\mathrm{d}}}{\Sigma_{\mathrm{g}}} \right)  \right) = \\
     = - \dot{\Sigma}_{\mathrm{d\_ext}},
\end{multline}
where $\Sigma_{\mathrm{d}}$ is the dust surface density, $D_{\mathrm{d}} = \nu/(1+\mathrm{St^{2}})$ is the diffusivity \citep{diffusion_coeff}, which depends on the kinematic viscosity $\nu$ and the Stokes number $\mathrm{St}$, and $\dot{\Sigma}_{\mathrm{d\_ext}}$ is the rate of dust entrained and lost in photoevaporative winds. The Stokes number $\mathrm{St}$ characterizes how much dust particles couple with the gas, and hence it determines the radial drift velocity. It is defined as the ratio of the dust stopping time and the the dynamical timescale, which in the Epstein regime (grain size $\ll$ particles mean free path) is given by \citep{dust_dynamics_2012}:
\begin{equation}
    \mathrm{St} = \frac{\pi}{2}\frac{a \rho_{\mathrm{m}}}{\Sigma_{\mathrm{g}}},
\end{equation}
where we are assuming spherical particles with radius $a$ and material density $\rho_{\mathrm{m}}$. At fixed dust and gas density, small grains (St$\ll$1) follow the gas motion, while large particles (St$\gg$1) move slowly in the radial direction and settle rapidly onto the mid-plane. At St$\sim$1 dust particles drift quickly towards the star following the pressure gradient (\citeauthor{Weidenschilling_77_drift} \citeyear{Weidenschilling_77_drift}; \citeauthor{whipple_72_drift} \citeyear{whipple_72_drift}).
\\

Dust growth through collisions is included in our model by solving the full coagulation equation known as Smoluchowski equation \citep{Smoluchowski_eq} for a distribution of particle masses.
Both radial drift and fragmentation are accounted for, considering that shorter drift timescales compared to growth timescales result in the removal of particles before they can achieve substantial growth. On the other hand, fragmentation occurs when dust grains collide at velocities exceeding their fragmentation threshold, $v_{\mathrm{frag}}$. This results in the outer regions being dominated by drift, while fragmentation becomes important at smaller radii \citep{dust_dynamics_2010}.
\\ 


As for the gas evolution model, we use the FRIEDv2 grid to estimate the rate of dust lost in photoevaporative winds, considering that dust particles smaller than a certain critical size, $a_{\mathrm{ent}}$, are entrained in the wind. The work of \cite{facchini_2016} demonstrates that this critical size is defined by the balance between the drag force exerted on dust grains by the wind and the gravity from the host star: 
\begin{equation}
    a_{\mathrm{ent}} = \frac{1}{4 \pi F} \frac{v_{\mathrm{th}} \dot{M}_{\mathrm{ext}}}{G M_{\star} \rho_{\mathrm{d}}},
\end{equation}
where $4 \pi F$ is the solid angle subtended by the wind, $v_{\mathrm{th}}$ is the velocity of the gas flowing outward, and $\rho_{\mathrm{d}}$ is the bulk density.
%
\begin{deluxetable}{cc}
    \tablecaption{Parameter space\label{tab:parameter_space}}
    \tablehead{
        \colhead{Parameter} & \colhead{Value}
        }
    \startdata
    $\alpha$ & $10^{-2}, \ 10^{-3}, \ 10^{-4}$ \\
    ${R_{\rm{c},0}}$ & [10, 40, 100] AU \\
    ${M_{\rm{disk},0}}$ & [0.01, 0.1] $\mathrm{M_{\star}}$ \\
    \enddata
    \tablecomments{Values for the $\alpha$ viscosity parameter, initial characteristic radius ${R_{\rm{c},0}}$, and disk mass $\mathrm{M_{disk,0}}$ explored in our simulations.}
\end{deluxetable}
%
%
\subsection{Simulation Setup}\label{subsec:simulation_setup}
Our numerical simulations are performed using \texttt{DustPy} \citep{dustpy} along with an integrated module that incorporates the effects of external photoevaporation \citep{Garate_2023} adapted to support the interpolation of the FRIEDv2 grid.
The gas surface density is initialized by the \citeauthor{LeP_solution} self-similar solution, considering a linear dependence of the viscosity from the radius $\nu \propto R^{\gamma}$ with $\gamma = 1$:
\begin{equation}
    \Sigma_{\mathrm{g}}(R) = \Sigma_{0}\left( \frac{R}{R_{\mathrm{c}}} \right)^{-1} \mathrm{exp}\left(-\frac{R}{R_{\mathrm{c}}}\right)
\end{equation}
where $R_{\mathrm{c}}$ is the characteristic radius, which corresponds to the exponential cut off and set the initial extent of the disk, and $\Sigma_{0}$ is set by the total disk mass.
  
The parameter space explored in our simulations is summarized in Table \ref{tab:parameter_space}. We studied disks with initial different masses and radial extent.
The values assumed by the $\alpha$ viscosity parameter are motivated by the empirical constraints required to explain disk accretion \citep{rosotti_turbolence}.
 
We assume that each target is subject to the FUV irradiation field evaluated as in Section \ref{sec: FUV calculation}. The temperature profile is based on the stellar luminosity (Eq. \ref{eq:temp_profile}) which, for the sake of simplicity and in consideration of the stellar masses of our targets, we set to a representative value of $0.3 \ L_{\odot}$.    

For all our simulations, we use a radial grid comprised of 250 cells equispaced in $R^{1/2}$.   
The radial extent of the grid is established based on the boundaries of the FRIEDv2 grid, with the inner and outer edge set to 5 AU and 500 AU respectively. We note that the grid outer edge is much larger than the observed extent of disks in Upper Sco ($\lesssim 200$ AU, \citeauthor{AGEPRO_IV_UpperSco} \citeyear{AGEPRO_IV_UpperSco})  

An important constraint is imposed by the minimum mass loss rate allowed by the FRIEDv2 grid. To ensure the reliability of our results, the mass loss rate must not drop below the lower limit set by the grid, which is $\sim 10^{-14} \ \rm{M}_{\odot} \ \rm{yr}^{-1}$. Therefore, we stop our simulations if the computed mass loss rate falls below this threshold. This approach has no influence on the final result since by the time the disk reaches the minimum mass loss rate, the disk is nearly completely dispersed.

    For the dust we consider a logarithmic grid in particle mass going from $10^{-12}$ g to $10^{5}$ g, corresponding to a grain size distribution covering from  0.5 $\rm{\mu}$m to 24 cm. The initial dust-to-gas ratio for the distribution of particles is 0.01 and the fragmentation velocity is set to $v_{\mathrm{frag}} = 10$ $\mathrm{m\ s^{-1}}$ (\citeauthor{dust_fragm} \citeyear{dust_fragm}, \citeauthor{dust_frag_1} \citeyear{dust_frag_1}, \citeauthor{dust_frag_2} \citeyear{dust_frag_2}).  
    \\

    In order to constrain the age up to which we should let the disks evolve in our simulations, we take into account the isochronal ages of the Upper Sco AGE-PRO sources derived from the stellar parameters \citep{AGEPRO_I_overview}, which spans between $\sim$2 and $\sim$6 Myr, on average (see Table \ref{tab:source_params}). Therefore, we compare the resulting disk masses and sizes at 2 and 6 Myr with the AGE-PRO observed values. Whenever the photoevaporative mass loss rate falls below the minimum allowed before 6 Myr, we stop the simulations at that time, as disks are nearly completely dispersed after that threshold and numerical results not reliable.

    \subsection{Evaluation of Disk Radius and Mass }\label{subsec:eval_radius_mass}

    We consider as gas disk radius from the AGE-PRO observations the radius in the disk enclosing the 90\% of $^{12}$CO line flux (see Table \ref{tab:source_params} and \citeauthor{AGEPRO_XI_gas_disk_sizes}\citeyear{AGEPRO_XI_gas_disk_sizes} for the derivation).
    This radius is a good approximation of the position in the disk where the $^{12}$CO column density drops below a critical threshold, inhibiting self-shielding and causing photo-dissociation. 
    In our models, this radius is evaluated considering that the gas column density at the radius enclosing the 90\% of $^{12}$CO line flux, $N_{\mathrm{gas}}(R_{^{12}\mathrm{CO}})$, which refers to column density of H$_{2}$, can be analytically prescribed as function of the disk mass and stellar luminosity \citep{Trapman_2023}:
    \begin{equation}
        N_{\mathrm{gas}}(R_{^{12}\mathrm{CO}, 90\%}) \approx 10^{21.27 - 0.53 \log(L_{\star})} \left( \frac{M_{\mathrm{gas}}}{M_{\odot}} \right)^{0.3 - 0.08 \log(L_{\star})},
        \label{n_crit}
    \end{equation}
    where $M_{\mathrm{gas}}$ is the gas mass of the disk at a fixed time, and the stellar luminosity is set to $0.3 L_{\odot}$ in our model. The above expression is retrieved from fitting the results of DALI thermochemical models and depends on the initial carbon abundance and the disk temperature.
    Given the critical column density, $R_{^{12}\mathrm{CO}}$ is computed in our model as the radius corresponding to the critical value of the disk surface density, at each time-step:
    \begin{equation}
        \Sigma(R_{^{12}\mathrm{CO}},t) = \Sigma_{\mathrm{crit}}(t) = N_{\mathrm{gas}}(R_{^{12}\mathrm{CO},90\%}, t) \mu_{\mathrm{gas}},
    \end{equation}
    where $\mu_{\mathrm{gas}}$ is the mean molecular weight of the gas in the disk, which we set to the standard value $\mu_{\mathrm{gas}} \sim 2.3 \ \mu_{\mathrm{H}} = 1.15 \ \mu_{\mathrm{H}_{2}}$.

    An external UV radiation heats the disk in its outer region, leading to an increase in temperature and subsequently altering the shape of the critical column density. In particular, as shown in \citet{Trapman_2023}, $N_{\mathrm{gas}}(R_{^{12}\mathrm{CO}, 90\%})$ increases in smaller disks ($\leq 10^{-4} M_{\odot}$) subject to a $30 \ \mathrm{G_{0}}$ FUV flux, gradually approaching the level observed in disks irradiated with 1 G$_0$ for larger masses.
    The expression for the critical column density shown in equation \eqref{n_crit} is valid when the external FUV flux is 1 G$_{0}$.
    We interpolated Figure 2 of \citet{Trapman_2023} to take into account the (constant) FUV flux applied.
    A higher $N_{\mathrm{gas}}$ due to higher FUV flux leads to higher $\Sigma_{\mathrm{crit}}$, which depending on the shape of the surface density can result in smaller or similar $R_{^{12}\mathrm{CO},90\%}$.
    \\
    
    For what concerns the dust component, from an observational point of view a measure of the disk mass is inferred from the sub-mm flux density (e.g. \citeauthor{Ansdell_2016} \citeyear{Ansdell_2016}, \citeauthor{Berenfeld_uppSco_2016} \citeyear{Berenfeld_uppSco_2016}). To compare models and observations, we compute the integrated flux density from the disk accounting for the contribution of all the dust grains included in the simulations at a given wavelength. If we neglect scattering and assume a face-on\footnote{The inclination do not influence the total flux as long as the disk is optically thin, but can lead to an overestimate of the flux (and the disk mass) if the disk is optically thick. Our models are always in the optically thin regime.}
    and vertical isothermal disk, the sub-mm flux density from the disk is 
    \begin{equation}
        F_{\nu} = \frac{1}{d^{2}} \int^{R_{\mathrm{out}}}_{R_{\mathrm{in}}} 2 \pi R B_{\nu}(T)(1- e^{-\tau_{\nu}}) dR,
        \label{eq:integrated_mm_flux}
    \end{equation}
    where $d$ is the distance to the source, which we consider $\sim$142 pc for Upper Sco, $B_{\nu}(T)$ is the Planck function at temperature $T$ and $\tau_{\nu} = \kappa_{\nu}(a) \Sigma_{\mathrm{dust}}$ is the vertical optical depth. The dust opacity $\kappa_{\nu}(a)$ is computed using the DSHARP Mie-Opacity Library \citep{Birnstiel_opacity_2018}, employing the volume fractions and the properties of the dust grains of the model by \citet{Ricci_2010}, for each particle size $a$. 
    The Planck function $B_{\nu}(T)$ is evaluated assuming a constant dust temperature $T = T_{\mathrm{dust}} = 20$ K, and all the dust properties are estimated at wavelength $1.3$ mm, which corresponds to the continuum Band 6 observations of AGE-PRO \citep{AGEPRO_IV_UpperSco}. 
    The flux estimate is obtained by taking into account that the sensitivity of the AGE-PRO survey is $\sim 3 \times 10^{7} \ \rm{Jy} \ \rm{sr}^{-1}$ \citep{AGEPRO_IV_UpperSco}.    
    The dust radius is determined as the position in the disk that encompasses 90\% of the continuum flux evaluated as in equation \eqref{eq:integrated_mm_flux}. 

\section{Simulation Results}\label{sec:results}
In this section, we present the outcomes of our simulations. In Section \ref{subsec:test}  we test how different FUV field intensities affect disk properties. Then, disk masses and sizes from our simulations are compared with those derived from the AGE-PRO measurements. In particular, the gas disk component is investigated in Section \ref{subsec:gas_models}, while in Section \ref{subsec:alpha} we discuss the role of the disk viscosity. 
In Section \ref{subsec:dust_models} we focus on the dust component, and in Section \ref{subsec:gas_to_dust_ratio} we investigate the gas-to-dust size ratio.
Finally, in Section \ref{subsec:varying_rc} we investigate how disk properties are affected by the initial disk extent when the disk evolution is driven by external photoevaporation and, in Section \ref{subsec:correlation} we discuss potential correlations between disk properties and FUV flux.

\subsection{Test the Effect of External Photoevaporation on Disk Properties}\label{subsec:test}

\begin{deluxetable}{cc}
    \tablecaption{Initial parameters of the test simulations\label{tab:inital_cond_example}}
    \tablehead{
        \colhead{Parameter} & \colhead{Value}
    }
    \startdata
    $M_{\star}$ & $0.6 \ \mathrm{M_{\odot}}$ \\
    $M_{\rm{disk},0}$ & 0.06 $\mathrm{M_{\odot}}$ \\
    ${R_{\rm{c},0}}$ & 10 AU \\
    $\alpha$ & $10^{-3}$ \\
    FUV & [0, 1, 10, 100, 1000] $\rm{G}_{0}$ \\
    \enddata
    
    \tablecomments{Stellar and disk parameters chosen for the test simulation. The FUV flux experienced varies from 0 to 1000 G$_{0}$.}
\end{deluxetable}

\begin{figure*} 
    \centering
    \includegraphics[width=0.7\textwidth]{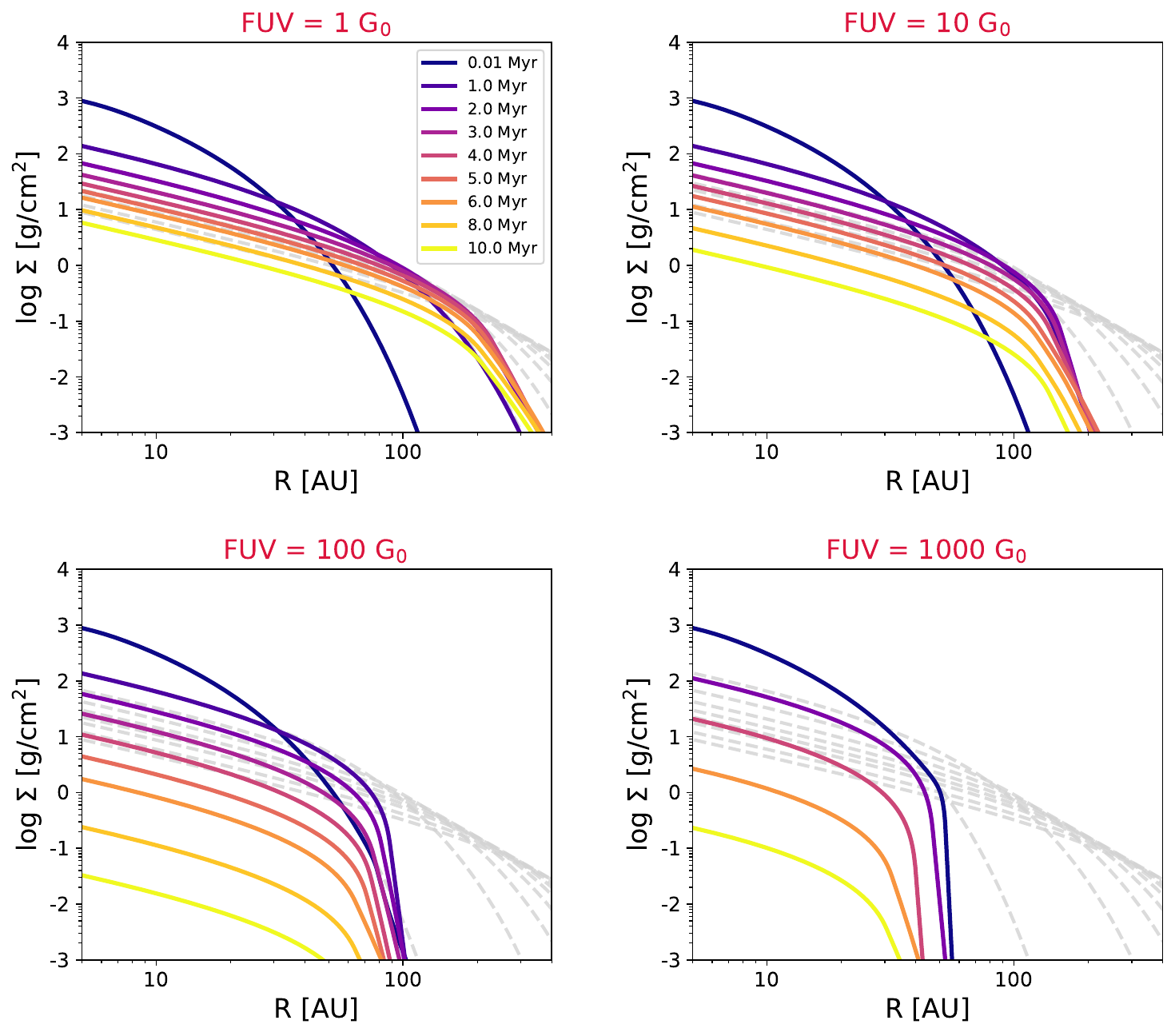}
    \caption{Evolution of the disk surface density when an external FUV field of 1, 10, 100 and 1000 G$_{0}$ is applied. Lines are color-coded based on the age, while the dashed gray lines corresponds to the 0 G$_{0}$ pure viscous evolution from 0 to 10 Myr. }
    \label{fig:sigma}
\end{figure*}

\begin{figure} 
    \centering
    \includegraphics[width=0.45\textwidth]{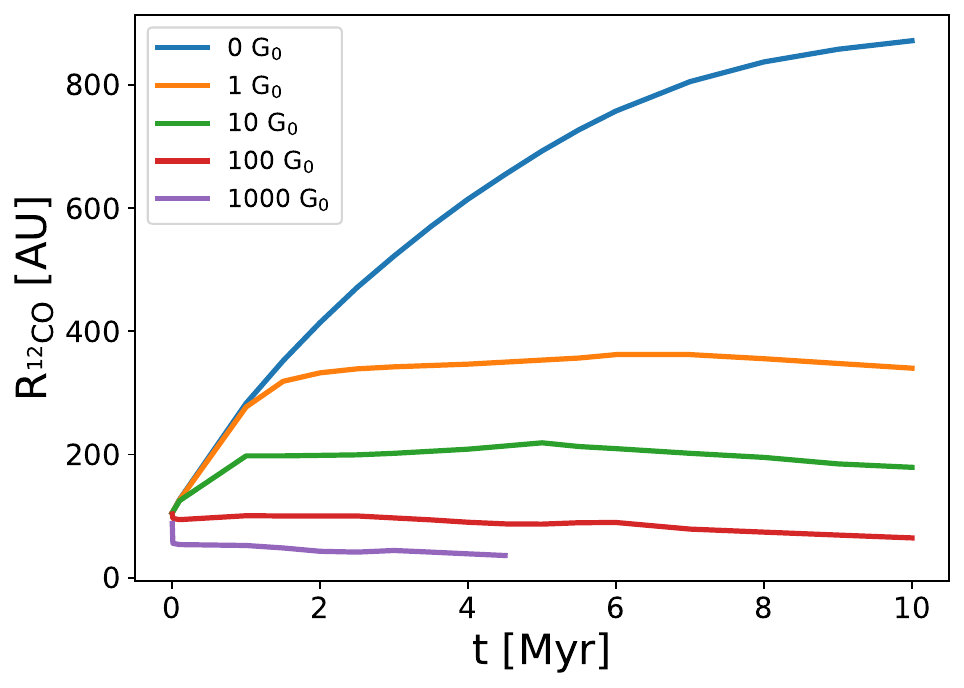}
    \caption{Time evolution of the disk radius when the FUV flux experienced is [0, 1, 10, 100, 1000]\,G$_{0}$. In the pure viscous model (0 G$_{0}$) the radius increases with time, but even low/moderate FUV fluxes (1-10 G$_{0}$) can halt viscous spreading. In the case of 1000 G$_{0}$, we stopped the simulation when the minimum mass loss rate allowed by the FRIEDv2 grid is reached, and the disk has almost completely dispersed in $\sim$5 Myr. }
    \label{fig:rgas_example}
\end{figure}

\begin{figure} 
    \centering
    \includegraphics[width=0.45\textwidth]{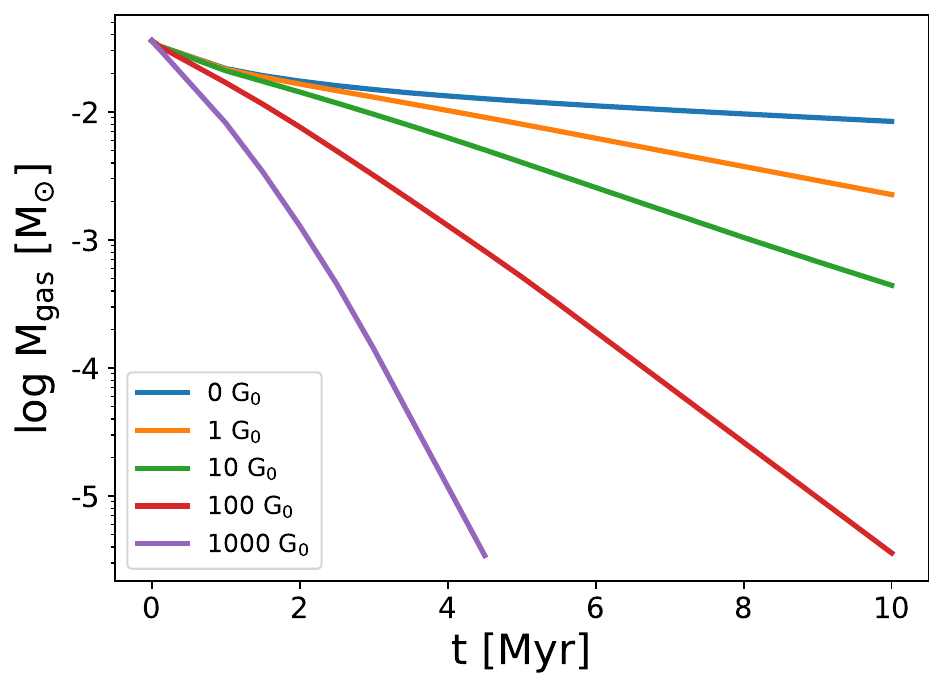}
    \caption{Time evolution of the total gas mass when the disk is subject to [0, 1, 100, 1000]\,G$_{0}$. }
    \label{fig:mgas_example}
\end{figure}

In order to explore the effect of external photoevaporation on disk mass and radius, we propose here an illustrative example in which we numerically evolve a protoplanetary disk, keeping the initial parameters constant except for varying the FUV flux. 
We use \texttt{DustPy} simulations and the initial parameters listed in Table \ref{tab:inital_cond_example}.
We let the simulations evolve until reaching 10 Myr, with the condition of stopping them earlier whether the photoevaporative mass loss drops below the minimum allowed by the FRIEDv2.

In Figure \ref{fig:sigma} we show in colored lines the time evolution of the gas surface density when the disk is subject to 1, 10, 100 and 1000 G$_{0}$, compared to the pure viscous framework (FUV = 0 G$_{0}$) in gray dashed lines. We observe that under low irradiation conditions (1 G$_{0}$ and 10 G$_{0}$), the early evolution of the disk, spanning the initial 2-3 million years, is primarily influenced by viscosity, resulting in spreading. However, as the mass loss rate becomes significant, due to the increase in disk size via viscous spreading, the subsequent evolution is driven by external photoevaporation, leading to disk shrinking.
Higher FUVs cause a more important loss of material in winds, and the disk is immediately truncated. In the 1000 G$_{0}$ scenario, the simulation stops before 5 Myr, when the mass loss rate reaches the FRIEDv2 limit, coinciding with the near-complete dispersal of the disk. 

It is crucial to point out that external photoevaporation is influencing disk evolution, leading to a substantial decrease in disk size, and in disk mass, even at low G$_{0}$, as shown in Figure \ref{fig:rgas_example} and \ref{fig:mgas_example}.
Under a low/moderate irradiation (1-10 G$_{0}$), disk mass at 10 Myr may decrease by one order of magnitude compared to the prediction of the pure viscous model. 
However, considering that through direct observations disk sizes are usually known with higher precision than disk masses, a slight decrease in size (of few AU) may be more evident than a greater decrease in mass (hundreds of $\mathrm{M}_{\odot}$).
This is evident even in the results of the high-resolution observations of AGE-PRO, where disk masses show large uncertainties compared to disk sizes (see Table \ref{tab:source_params} and the following Figure \ref{fig:mass_gas} and \ref{fig:rco}).
Furthermore, in the range [1-10] G$_{0}$, disk sizes are significantly reduced by the first 2-4 Myr, whereas the difference in disk mass becomes relevant after 6-8 Myr. Therefore, particular attention should be directed towards size-related studies, especially if we want to investigate the role of external photoevaporation in moderately irradiated environments such as Upper Sco.
We highlight that extreme caution should be applied if no viscous spreading is observed in mildly irradiated regions, as our results show that even moderate levels of irradiation can suppress disk spreading.

Since the FUV fluxes for the AGE-PRO targets in Upper Sco span (on average) from 2 to 12 G$_{0}$ (Figure \ref{fig:fuv_flux}), the results presented above motivate our investigation into disk masses and sizes accounting for external photoevaporation. 
We emphasize that the consequences on the overall disk evolution will be more or less significant depending on the stellar mass and initial disk parameters. These dependencies are explored in the following sections.   
    \subsection{AGE-PRO Gas Results}\label{subsec:gas_models}

    \begin{figure}[htbp]
    \centering
    \begin{minipage}[b]{0.45\textwidth}
        \centering
        \includegraphics[width=\textwidth]{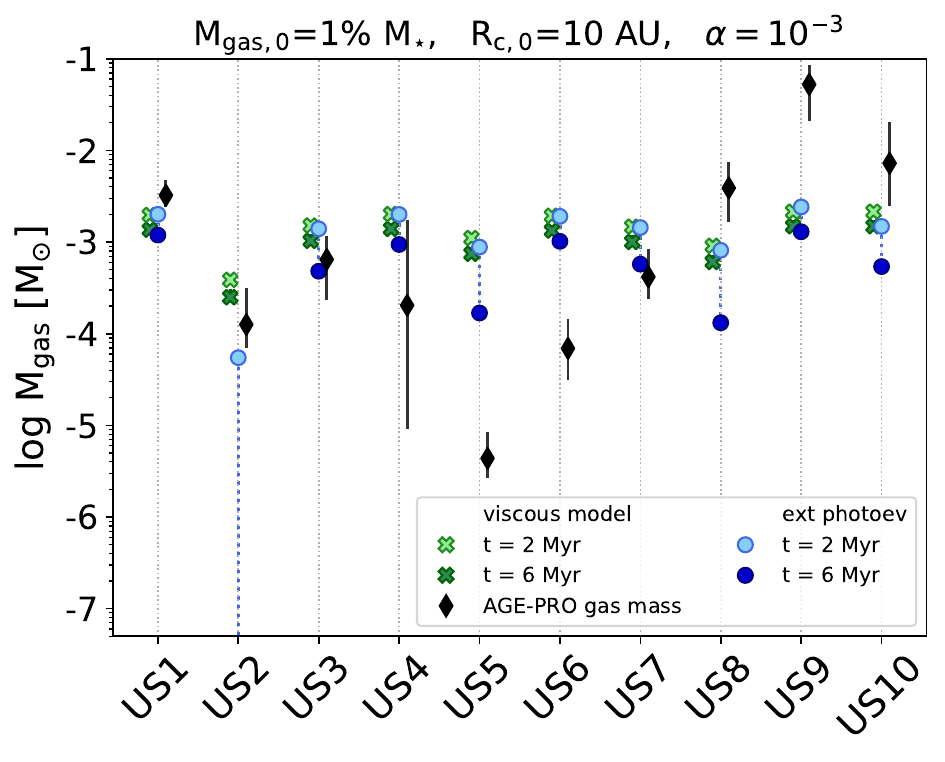}
    \end{minipage}
    \hfill
    \hspace{0.002\textwidth}
    \begin{minipage}[b]{0.45\textwidth}
        \centering
        \includegraphics[width=\textwidth]{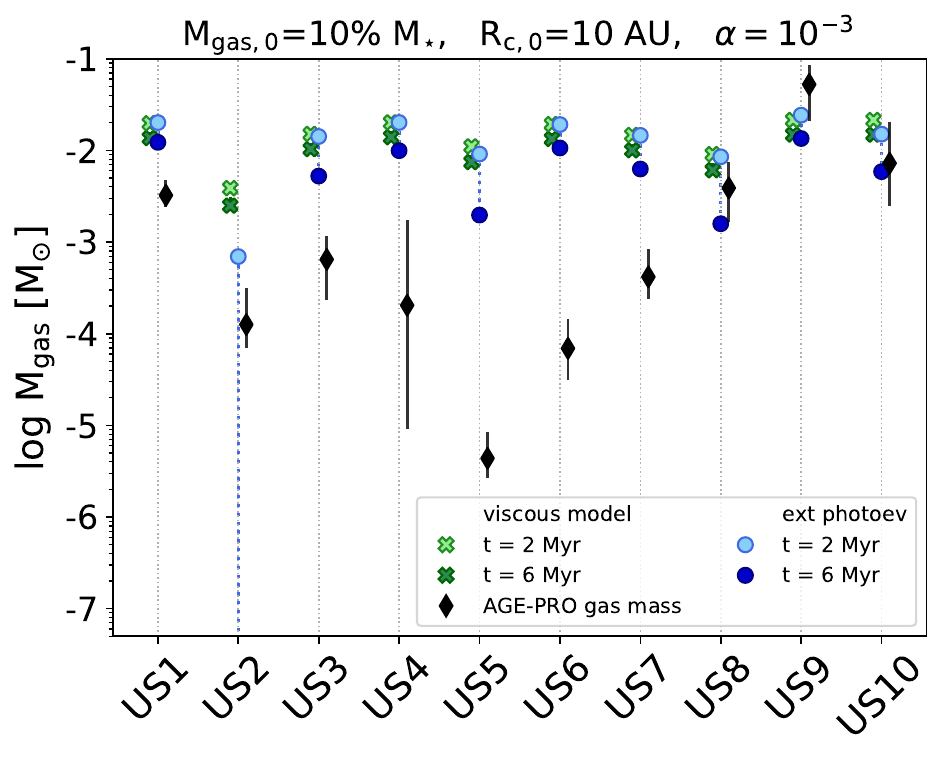}
    \end{minipage}
    \caption{Gas disk masses in the purely viscous scenario (green crosses) and adding the effect of external photoevaporation (blue dots), compared with the gas mass from AGE-PRO (black diamonds). \emph{Top}: Initial disk mass set to 1\% $\rm{M}_{\star}$. \emph{Bottom}: Initial disk mass set to 10\% $\rm{M}_{\star}$. The dotted line associated with UppSco 2 is not connected to any blue dot, as it is predicted to be almost fully dispersed due to photoevaporation by 6 Myr.}
    \label{fig:mass_gas}
    \end{figure}
    \begin{figure}[htbp]
    \centering
    \includegraphics[width=0.45\textwidth]{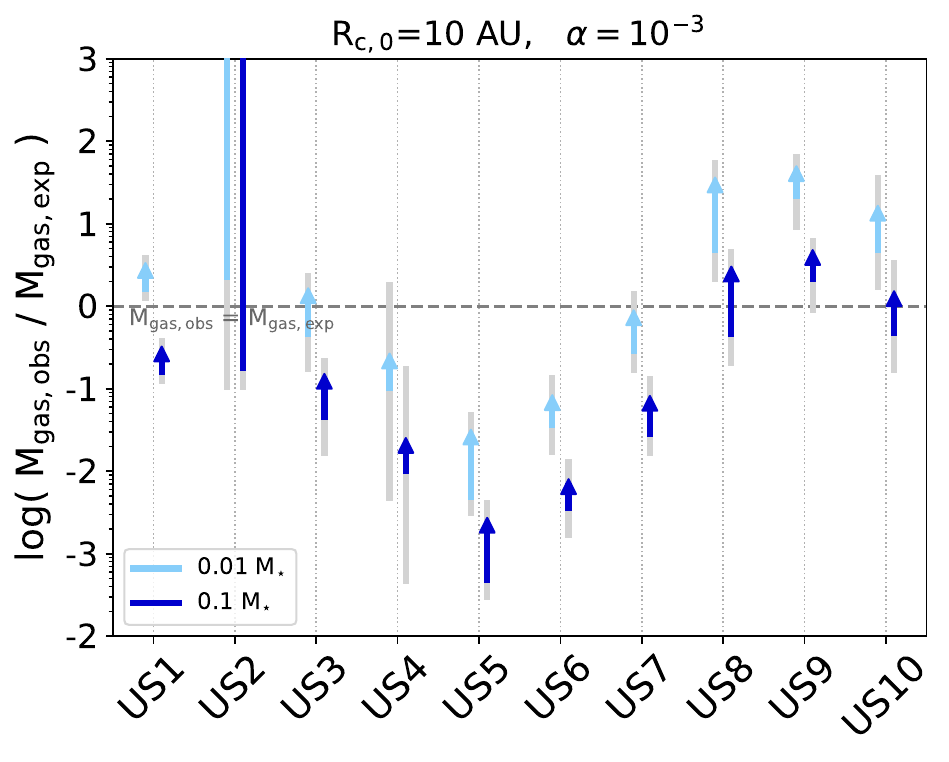}
    \caption{Ratio between the observed gas masses and the those predicted when external photoevaporation is included in the model. The colored lines refer to the values in the range of ages [2 - 6] Myr (with arrows pointing from younger to older ages) considering an initial disk mass of 0.01 M$_{\star}$ and 0.1 M$_{\star}$, respectively shown in light blue and dark blue. The shaded gray regions refer to the uncertainties in the observed gas masses.}
    \label{fig:diff_mass_gas}
    \end{figure}
    \begin{figure}[htbp]
    \centering
    \begin{minipage}[b]{0.45\textwidth}
        \centering
        \includegraphics[width=\textwidth]{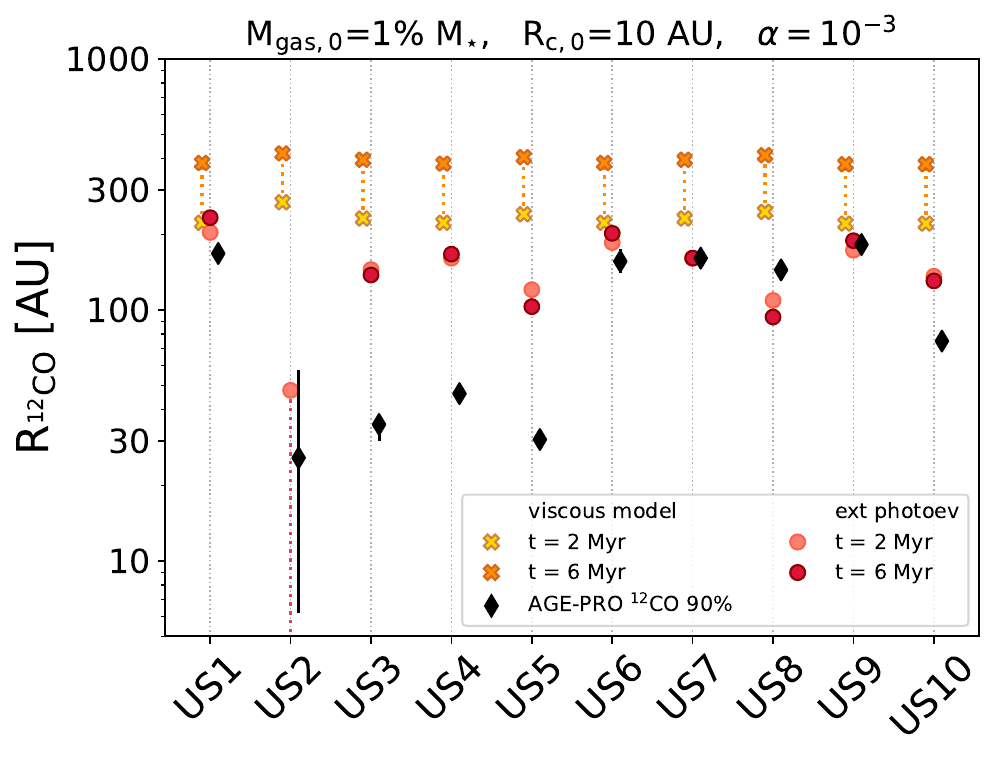}
    \end{minipage}
    \hfill
    \hspace{0.002\textwidth}
    \begin{minipage}[b]{0.45\textwidth}
        \centering
        \includegraphics[width=\textwidth]{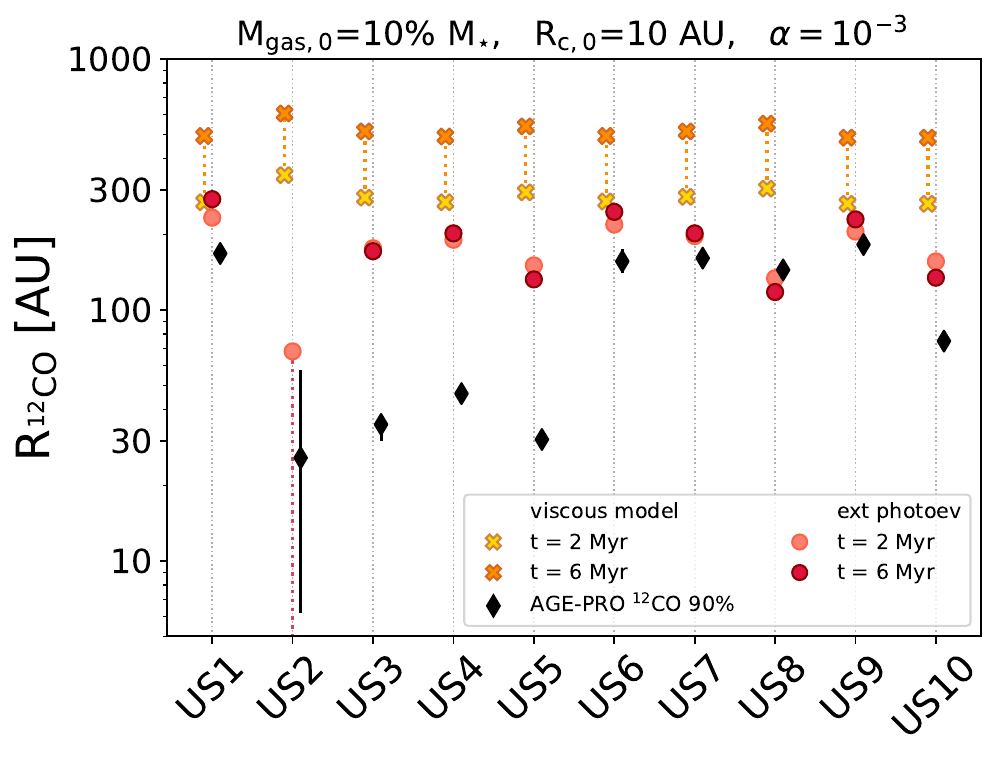}
    \end{minipage}
    \caption{$\mathrm{R}_{^{12}\mathrm{CO}}$ from our simulations in the purely viscous scenario (orange crosses) and adding external photoevaporation (red dots). Comparison with the radius enclosing the 90\% of the $^{12}\mathrm{CO}$ flux, observed by AGE-PRO. \emph{Top} and \emph{Bottom} panels refer to models with initial disk mass set to 1\% $\rm{M}_{\star}$ and 10\% $\rm{M}_{\star}$, respectively. The line related to UppSco 2 is not connected to any red dot, as it is predicted to be almost fully dispersed due to photoevaporation in $<$ 6 Myr.}
    \label{fig:rco}
    \end{figure}
    Focusing for the moment on the gas disk component, we study the evolution of the ten disks in both the purely viscous scenario and when subject to an external FUV radiation field.
    In this Section we provide the results of the models in which we set the initial characteristic radius, $R_{\rm{c},0}$, to 10 AU, $\alpha = 10^{-3}$, and the initial disk mass to either 1\% and 10\% of the stellar mass, for each source.
    
    In Figure \ref{fig:mass_gas} the disk mass values predicted by our models, at 2 Myr and 6 Myr, are compared with the disk mass retrieved from the AGE-PRO observations \citep{AGEPRO_V_gas_masses}.
    The inclusion of an external radiation field causes a reduction in disk mass compared to the viscous framework, which becomes more evident at older ages (see green and blue markers in Figure \ref{fig:mass_gas}).
    The observed gas masses of eight out of ten targets are consistent within the error bars with the results of the model incorporating external photoevaporation, if the initial disk mass is between 1\% and 10\% of the stellar mass, within a factor of 2 (see Figure \ref{fig:diff_mass_gas}).
    UppSco 5 and UppSco 6 are closer to the predictions made by the model with initial disk mass 0.01 M$_{\star}$, but too low-mass to be well represented by the model used, as shown in Figure \ref{fig:diff_mass_gas}.
    Among the other eight targets, four are represented better by the case where the initial disk mass is 0.01 $\rm{M}_{\star}$ (UppSco 1, UppSco 3, UppSco 4, and UppSco 7, as shown in the top panel of Figure \ref{fig:mass_gas} and in Figure \ref{fig:diff_mass_gas}), and four sources (UppSco 2, UppSco 8, UppSco 9, and UppSco 10) prefer an higher initial disk mass of 0.1 $\rm{M}_{\star}$ (see the bottom panel in Figure \ref{fig:mass_gas} and Figure \ref{fig:diff_mass_gas}).
    However, relying solely on the analysis of the gas mass is insufficient for quantifying the impact of the external photoevaporation, as well as constraining the initial conditions. Therefore, in this study we intend to correlate this result with the information provided by gas radii and dust properties to derive our conclusions.
    \\
    
    UppSco 2 is predicted to be almost entirely dispersed
    in less than 6 Myr due to external photoevaporation, considering it to be exposed to $\sim 12 \ \rm{G}_{0}$. This result remains unchanged even when considering an initial higher mass for the disk (0.1 M$_{\star}$ in our case)     
    and is caused by the combination of a small stellar mass ($\sim$0.13 $\mathrm{M}_{\odot}$, Table \ref{tab:source_params}) against an experienced FUV flux of $\sim 12 \ \rm{G}_{0}$ (Fig. \ref{fig:fuv_flux}). 
    UppSco 2 is the only disk for which we stopped the stimulation including external photoevaporation before reaching 6 Myr.
    The actual detectability of UppSco 2 may be attributed to a lower FUV exposure than what we assumed. Notably, we evaluated a significant uncertainty in the external flux experienced by this target, as shown in Figure \ref{fig:fuv_flux}. Moreover, as often discussed in the context of proplyds, the dynamics of the cluster can alter the relative distances between disks and massive stars, leading to variations in the FUV flux with time. Additional discussion on this topic will be carried out in Section \ref{sec:discussion}. 
    \\

    In Figure \ref{fig:rco}, the disk radii resulting from our simulations are compared with the AGE-PRO $R_{^{12}\mathrm{CO}, 90\%}$ retrieved from the data analysis \citep{AGEPRO_XI_gas_disk_sizes}.
    As highlighted in previous works \citep{Trapman_2020}, the observed radii (black diamonds in Figure \ref{fig:rco}) are highly inconsistent with viscous models (orange and yellow dots in Figure \ref{fig:rco}). The disparity between model and observations significantly decreases with the inclusion of the external photoevaporation.
    This result definitively shows the importance of accounting for the process of external photoevaporation even in regions where the FUV radiation field is moderate, as observed in Upper Sco.
    
    While the inclusion of photoevaporation drastically reduces gas disk sizes, not all our targets shrink over time (within the time range considered): between 2 and 6 Myr, UppSco 1, UppSco 6, and UppSco 9 are still predicted to experience viscous spreading, although the observed spreading is less significant compared to the pure viscous scenario. 
    
    The observed disk sizes of UppSco 3, UppSco 4, UppSco 5, and to a lesser extent, UppSco 10 are even smaller than predicted when accounting for external photoevaporation. This discrepancy might be caused by additional processes contributing to further disk depletion (e.g. MHD-driven winds, which could be relevant especially for UppSco 3 and UppSco 10, \citeauthor{Fang_pascucci_2023} \citeyear{Fang_pascucci_2023}), as well as a higher FUV flux than initially estimated, as we discuss in Section \ref{subsec: discussion_additional_factors}. Moreover, tidal truncation produced by the presence of a stellar or planetary companion results in a smaller disk radius. 
\subsection{Varying the disk viscosity}\label{subsec:alpha}
\begin{figure}[htbp]
    \centering
    \begin{minipage}[b]{0.45\textwidth}
        \centering
        \includegraphics[width=\textwidth]{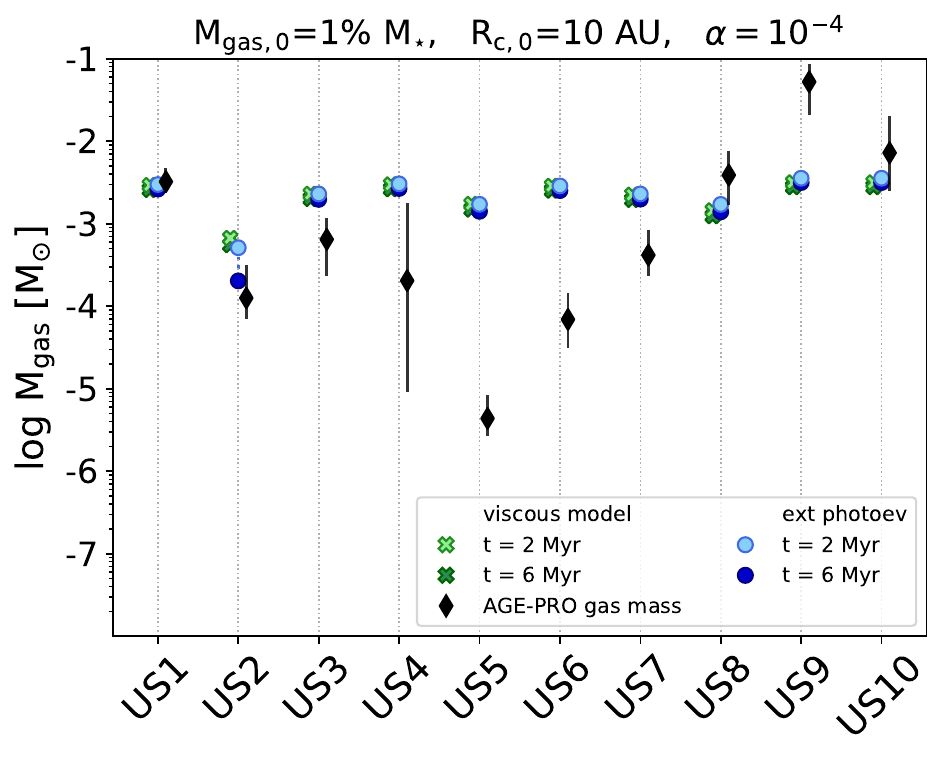}
    \end{minipage}
    \hfill
    \hspace{0.002\textwidth}
    \begin{minipage}[b]{0.45\textwidth}
        \centering
        \includegraphics[width=\textwidth]{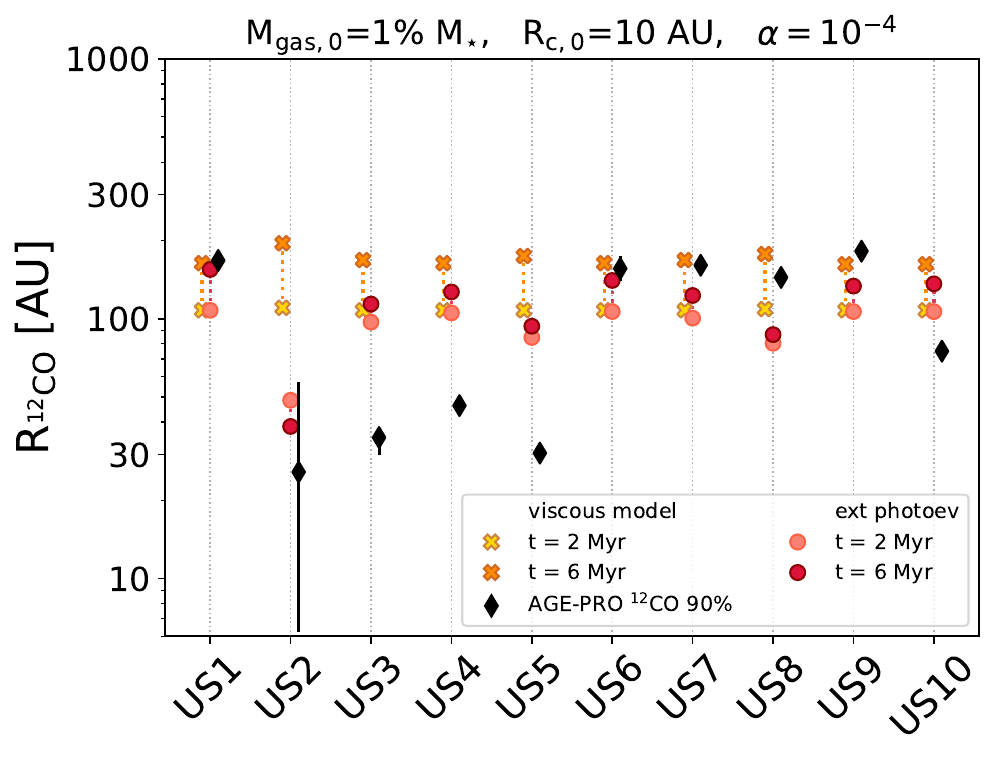}
    \end{minipage}
    \caption{Gas disk masses (\textit{top panel}) and radii (\textit{bottom panel}) resulting from models where $\alpha = 10^{-4}$. The results from using the viscous model and adding external photoevaporation are compared to the AGE-PRO observations. }
    \label{fig:alpha_4}
    \end{figure}
\begin{figure}[htbp]
    \centering
    \begin{minipage}[b]{0.45\textwidth}
        \centering
        \includegraphics[width=\textwidth]{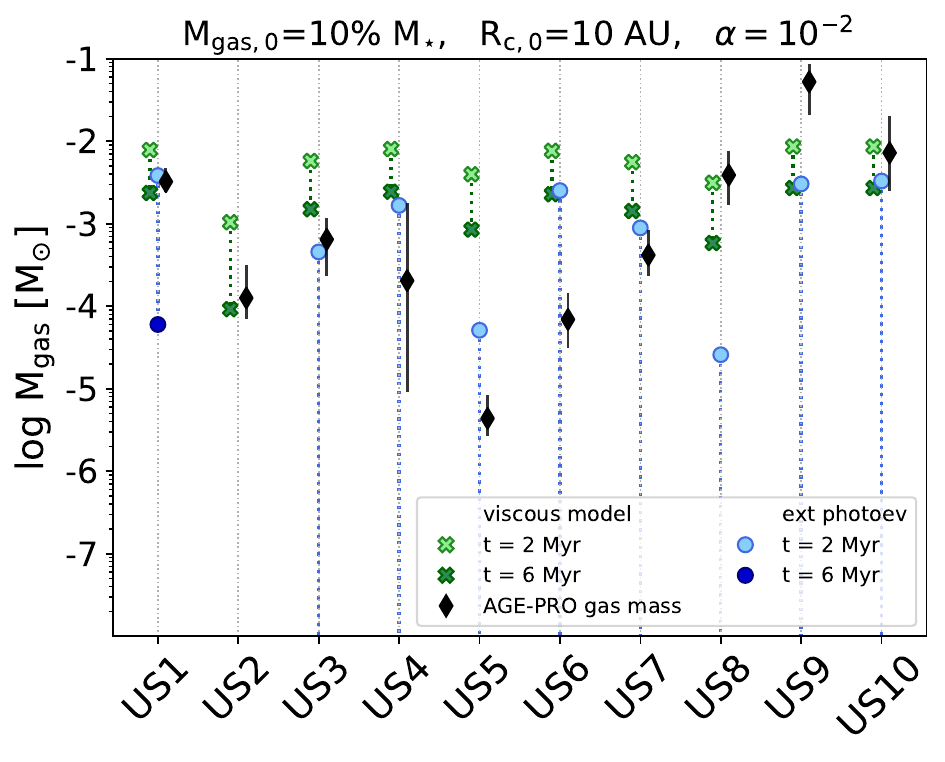}
    \end{minipage}
    \hfill
    \hspace{0.002\textwidth}
    \begin{minipage}[b]{0.45\textwidth}
        \centering
        \includegraphics[width=\textwidth]{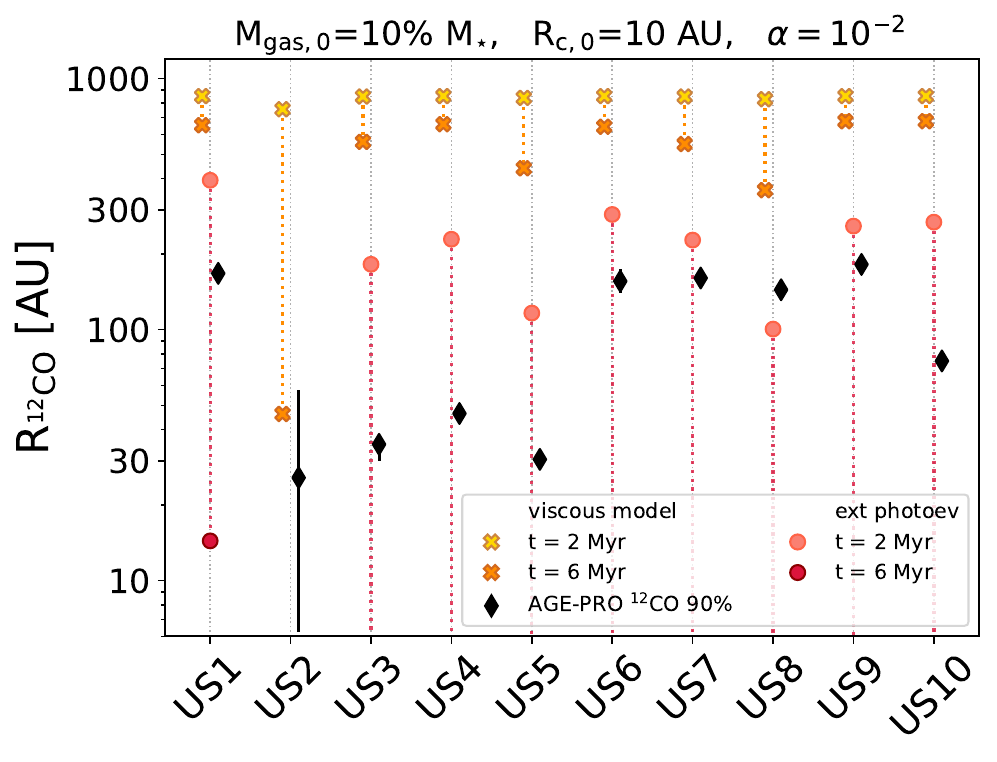}
    \end{minipage}
    \caption{Gas disk masses (\textit{top panel}) and radii (\textit{bottom panel}) considering $\alpha = 10^{-2}$ in our simulations. We compare the results from the viscous model with those including external photoevaporation and AGE-PRO observations. Dotted lines without end point means that disks are dispersed in less than 6 Myr.}
    \label{fig:alpha_2}
    \end{figure}
In this Section we discuss the impact of varying the $\alpha$ viscosity parameter, studying disk evolution when $\alpha=10^{-4}$ and $\alpha=10^{-2}$.
The comparison between models with $\alpha=10^{-4}$ and $\alpha=10^{-2}$, both with and without external photoevaporation, and the observed masses and radii from AGE-PRO is presented in Figure \ref{fig:alpha_4} and \ref{fig:alpha_2}. We show only the case in which $M_{\rm{gas},0} = 1\% \ \mathrm{M}_{\star}$ when $\alpha=10^{-4}$, and $M_{\rm{gas},0} = 10\% \ \mathrm{M}_{\star}$ when $\alpha=10^{-4}$, as the other models disagree more from observations.

According to the viscous theory, a higher viscosity results in a greater mass accretion rate, implying a shorter disk evolution timescale. As a consequence, after the same amount of time the disk experiences a more significant mass reduction and spreads more efficiently.
Consequently, the effect of external photoevaporation is expected to be more relevant if the disk is more viscous.
However, relying solely on the analysis of the resulting gas radii, and comparing them with the observations (Figure \ref{fig:alpha_4}) is insufficient to establish a preference between $\alpha = 10^{-4}$ and $\alpha = 10^{-3}$. While UppSco 3, UppSco 4, and UppSco 5 still present too small gas disk sizes to be explained by the inclusion of photoevaporation, the other targets show gas sizes that could be consistent with the models.
However, the large discrepancies in disk mass (see Figure \ref{fig:alpha_4}) disfavour the model with $\alpha = 10^{-4}$ for the majority of the sources.  
An exception is UppSco 2, if we assume an the initial disk mass of 0.01 M$_{\star}$. We remark that due to the substantial uncertainties in the FUV flux experienced by this target, it is challenging to make reliable claims.

In the case of $\alpha = 10^{-2}$ (Figure 
\ref{fig:alpha_2}), the viscous model completely fails in reproducing gas disk sizes and the majority of gas disk masses. When external photoevaporation is included, UppSco 2 is dispersed in 1 Myr, UppSco 5 and UppSco 8 in 3.5 Myr, while the other sources are dispersed between 5 and 6 Myr.
The sole survivor is UppSco 1 if $M_{\mathrm{gas,0}}=0.1 \ \mathrm{M}_{\odot}$.
The observed gas mass and size of UppSco1, UppSco 4, UppSco 5, UppSco 6, and UppSco 7 may be well reproduced by this model if the age of the those sources is bigger than 2 Myr but smaller than 6 Myr (the other sources present inconsistency between masses and sizes).
Given the large uncertainties in the isocronal age estimate of the targets (see Table \ref{tab:source_params}), only UppSco 4 and UppSco 7 present a range of ages covering values smaller than 6 Myr that make them potentially consistent with $\alpha=10^{-2}$. 
Assuming UppSco 5 and UppSco 6 to be younger than 6 Myr, the model with $\alpha=10^{-2}$ may explain the observed properties better than the $\alpha = 10^{-3}$ model. However, the large error bars in disk age prevent us from favouring $\alpha = 10^{-2}$ against $\alpha = 10^{-3}$ for these two sources.

We highlight that the claim that external photoevaporation is affecting disk masses and sizes in Upper Sco disks is supported by AGE-PRO observations even when a higher $\alpha$ viscosity is assumed.\\

As final remark, it is worth noticing that the $\alpha$ alpha is expected to vary over time and across the disk \citep{Delage_2022}, while we assume it as a constant. Given our incomplete understanding of disk accretion mechanisms, we can consider the selected value as a representative average across the disk extent and lifetime.
\\

\subsection{AGE-PRO Dust Results}\label{subsec:dust_models}

    \begin{figure}[htbp]
    \centering
    \begin{minipage}[b]{0.45\textwidth}
        \centering
        \includegraphics[width=\textwidth]{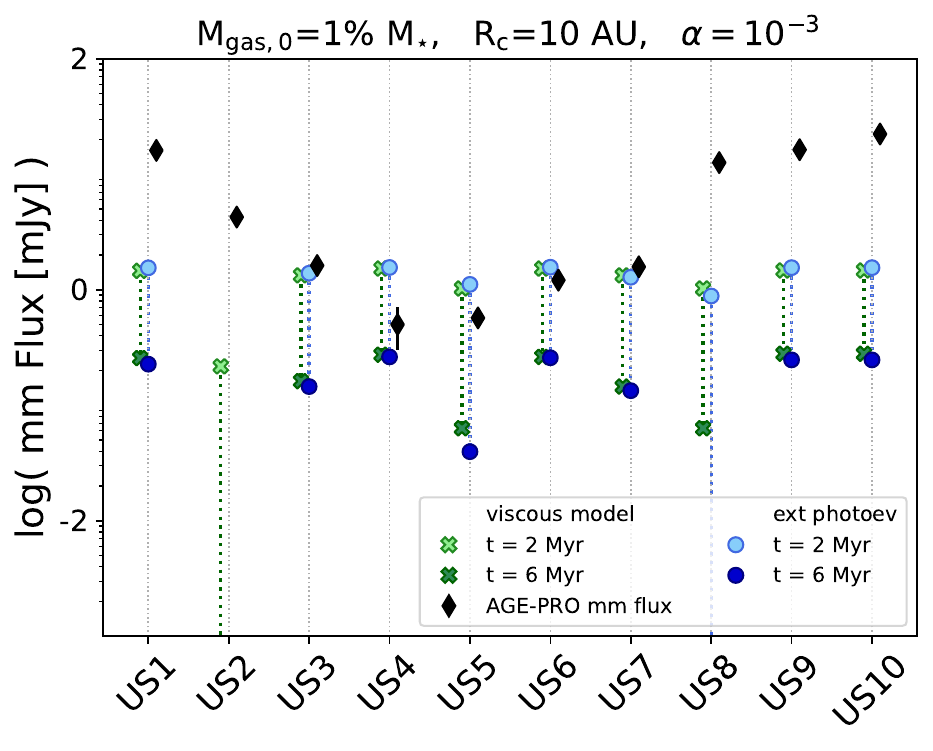}
    \end{minipage}
    \hfill
    \hspace{0.002\textwidth}
    \begin{minipage}[b]{0.45\textwidth}
        \centering
        \includegraphics[width=\textwidth]{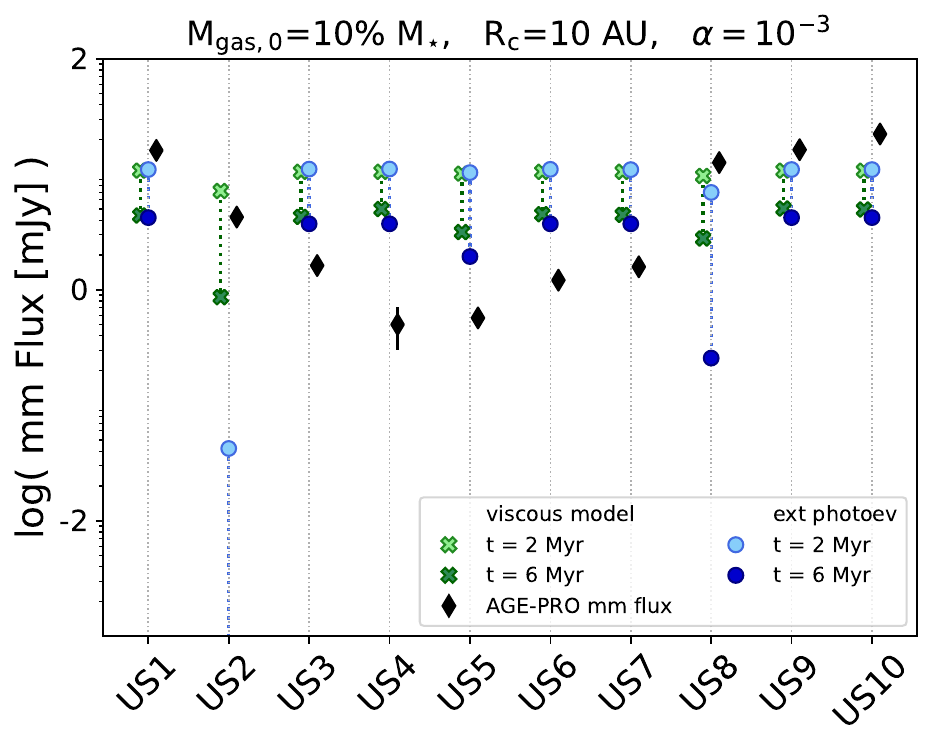}
    \end{minipage}    
    \caption{Comparison between the observed 1.3 mm continuum flux density of the ten AGE-PRO targets and the simulation results at 2 and 6 Myr, considering the purely viscous model (green crosses) and adding external photoevaporation (blue dots). The panels show the outcomes considering an initial disk mass of 1\%  and 10\% of the stellar mass, respectively shown on \emph{top} and \emph{bottom} panel.}
    \label{fig:mm_flux}
    \end{figure}

    \begin{figure}[htbp]
    \centering
    \begin{minipage}[b]{0.45\textwidth}
        \centering
        \includegraphics[width=\textwidth]{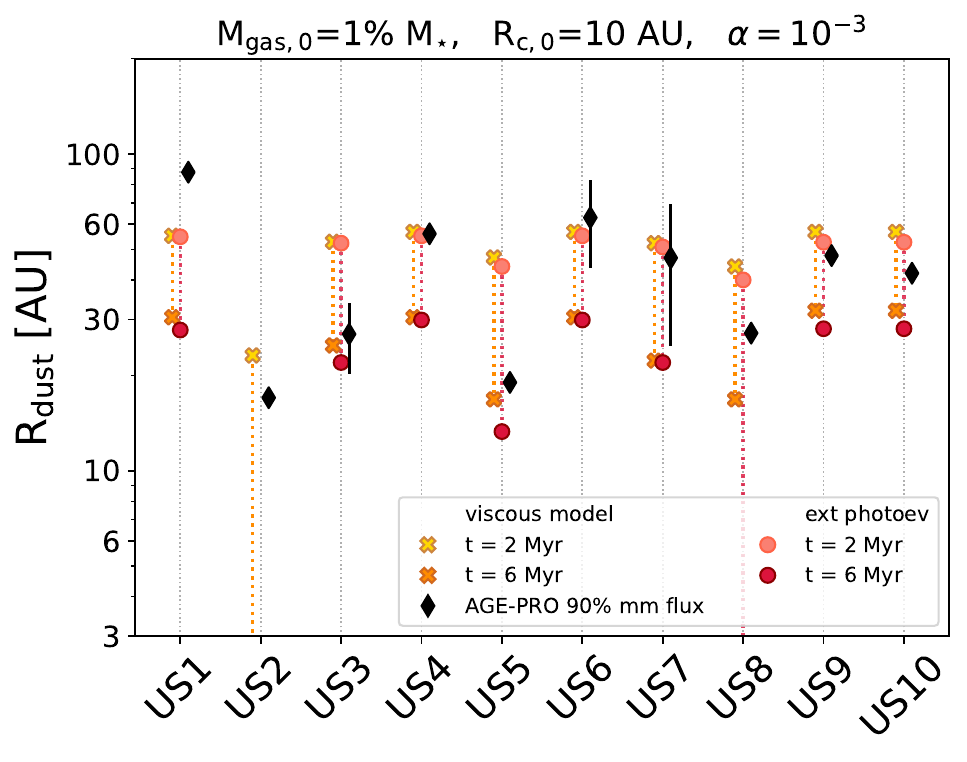}
    \end{minipage}
    \hfill
    \hspace{0.002\textwidth}
    \begin{minipage}[b]{0.45\textwidth}
        \centering
        \includegraphics[width=\textwidth]{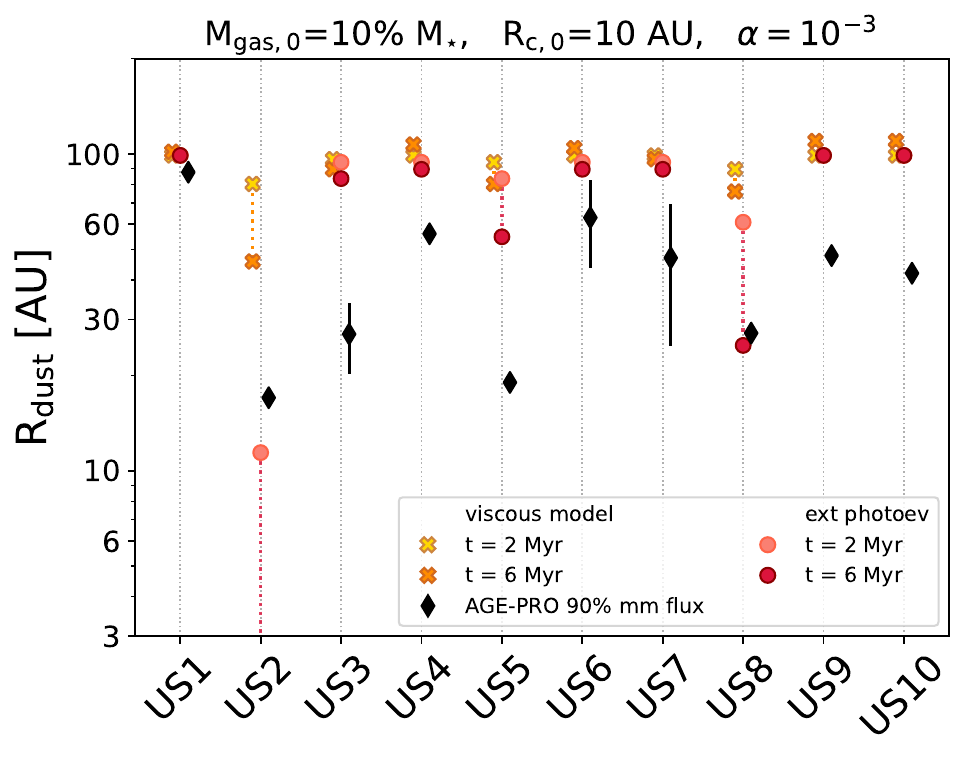}
    \end{minipage} 
    \caption{Dust radius: enclosing the 90\% of the 1.3 mm continuum flux, determined through AGE-PRO visibilities (black diamonds, where the values for UppSco 2, UppSco 4 and UppSco 5 are upper limits), and from our simulation with and without the effect of external photoevaporation (in orange crosses and red dots). \textit{Top} and \textit{bottom} panels show the results for an initial disk mass of 1\% and 10\% of the stellar mass, respectively. }
    \label{fig:rdust}
    \end{figure}

    In Figure \ref{fig:mm_flux} we compare the integrated emission at 1.3 mm observed by AGE-PRO with the integrated flux resulting from our simulations, considering both scenarios with and without the contribution of the external FUV field. 
    In the case of an initial disk mass set to 0.01 $\rm{M}_{\star}$, the targets exhibit millimeter fluxes above (or slightly consistent with) the model's predictions, even when considering the purely viscous framework. A similar pattern is observed for the dust radii, as shown in Figure \ref{fig:rdust}.    
    Increasing the initial disk mass leads to an overestimate of the dust disk sizes, with few sources being slightly consistent with the external photoevaporation model.
    In our models we adopted as dust radius the position in the disk enclosing the 90\% of the millimeter continuum flux, evaluated using equation \eqref{eq:integrated_mm_flux}. This analytic value is compared with the corresponding $R_{\mathrm{dust},90\%}$ retrieved from the AGE-PRO visibility fitting of continuum Band 6 observations \citep{AGEPRO_X_dust_disks}.
    
    In general, the models are not able to reproduce consistently both dust radii and mm fluxes. A potential solution to this is the role of dust substructures  \citep{AGEPRO_dust_evolution_Nico}, which are not included in our models.
    \subsection{Gas and dust relative sizes }\label{subsec:gas_to_dust_ratio}
    \begin{figure}[htbp]
    \centering
    \begin{minipage}[b]{0.45\textwidth}
        \centering
        \includegraphics[width=\textwidth]{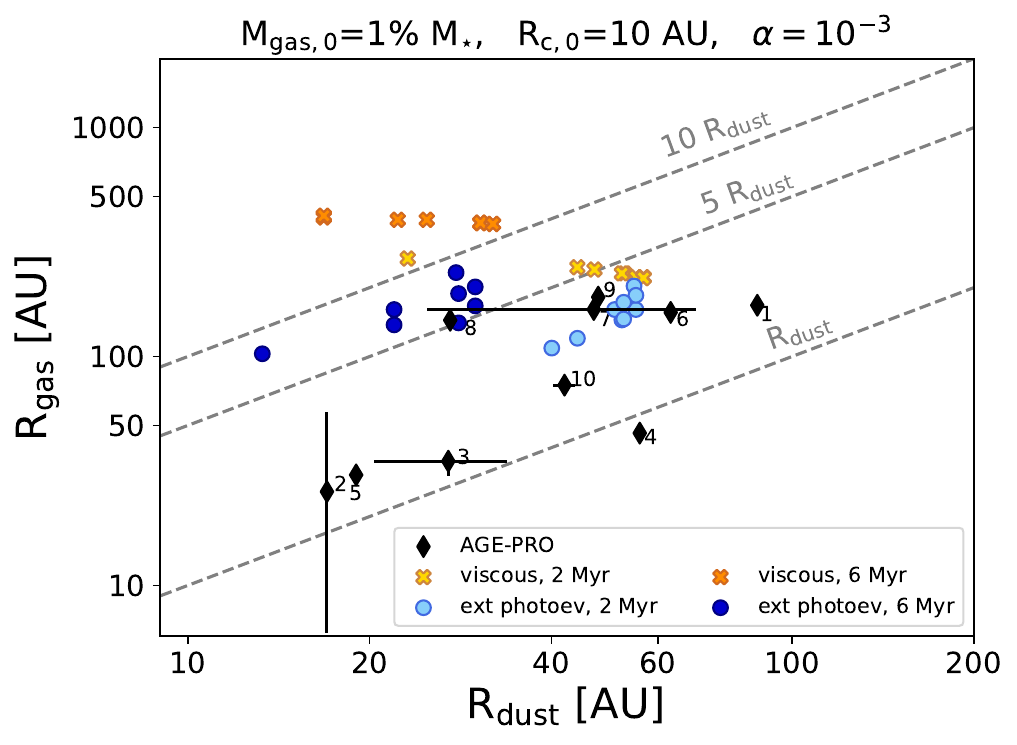}
    \end{minipage}
    \hfill
    \hspace{0.002\textwidth}
    \begin{minipage}[b]{0.45\textwidth}
        \centering
        \includegraphics[width=\textwidth]{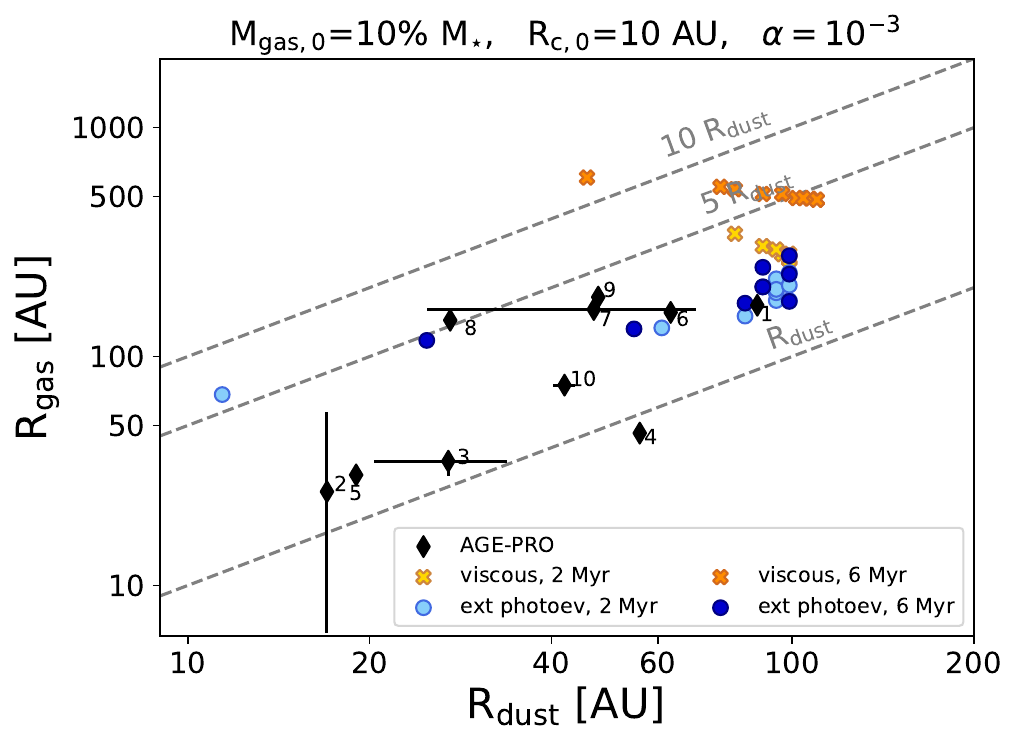}
    \end{minipage}
    \caption{The gas and dust sizes detected by AGE-PRO in Upper Sco (black diamonds) are compared with the simulations results after 2 Myr (in light colors) and 6 Myr (in dark colors). The simulations refer to an initial disk mass of 1\% and 10\% of the stellar mass on the top and bottom panels, respectively. Orange and yellow dots are related to the sole viscous evolution model, while blue and light blue dots refer to the models where external photoevaporation is also included.}
    \label{fig:rco_rdust}
    \end{figure}
    Gas and dust relative sizes resulting from our models are compared with the AGE-PRO observations. 
    For the bulk population of disks in non-irradiated star-forming regions, it is known that smooth model fails in reproducing the observed gas-to-dust relative sizes \citep{Toci_2021}, specifically producing a dust disk that is too small for a given gas disk. Moreover, the size ratio observed in Upper Sco is not too different from Lupus (\citeauthor{Sanchis_Testi_Lupus} \citeyear{Sanchis_Testi_Lupus}, \citeauthor{AGEPRO_I_overview} \citeyear{AGEPRO_I_overview}). Because we showed that gas disk sizes are reduced by external photoevaporation, it is worth investigating whether our models yield similar conclusions. We verify that this discrepancy holds in Upper Sco as well, as can be seen in Figure \ref{fig:rco_rdust}.
    
    The viscous (smooth) model predicts higher gas and dust relative sizes (identified as orange and yellow dots in Figure \ref{fig:rco_rdust}) with respect to the AGE-PRO observations (black diamonds in Figure \ref{fig:rco_rdust}), as already found by other authors (e.g. \citeauthor{Toci_2021} \citeyear{Toci_2021}).
    If an external FUV field is experienced, disks are significantly truncated in the gas (see Figure \ref{fig:rco}). However, dust disk sizes are also expected to decrease with time, due to both photoevaporation and radial drift. While the gas-to-dust size ratio does decrease and bring our models closer to the observations, the decrease is not significant enough to solve the discrepancy, which needs a different explanation.
    
    Considering that the AGE-PRO targets in Upper Sco exhibit gas sizes ranging from one to five times larger than the corresponding dust sizes,
    the results shown in Figure \ref{fig:rco_rdust} suggest that the sources are better represented by the scenario that includes external photoevaporation and $M_{\rm{disk},0} = 0.1 \ \rm{M_{\star}}$. However, this apparent correlation with an high initial disk mass is primarily due to the models inability to reproduce dust radii (see Figure \ref{fig:rdust}). Indeed, the presence of substructures is known to slow down the dust inward drift, leading to changes in the disk size (\citeauthor{Stadler_2022_substructures} \citeyear{Stadler_2022_substructures}, \citeauthor{Toci_2021} \citeyear{Toci_2021}). This process persists also when external photoevaporation is included \citep{Garate_2023}, highlighting the necessity of substructures to account for the observed gas-to-dust size ratios.

\subsection{ Varying the Initial Disk Extent:
    \\
    UppSco 6 and UppSco 8}\label{subsec:varying_rc}
    \begin{figure*}
        \centering
        \includegraphics[width=1\linewidth]{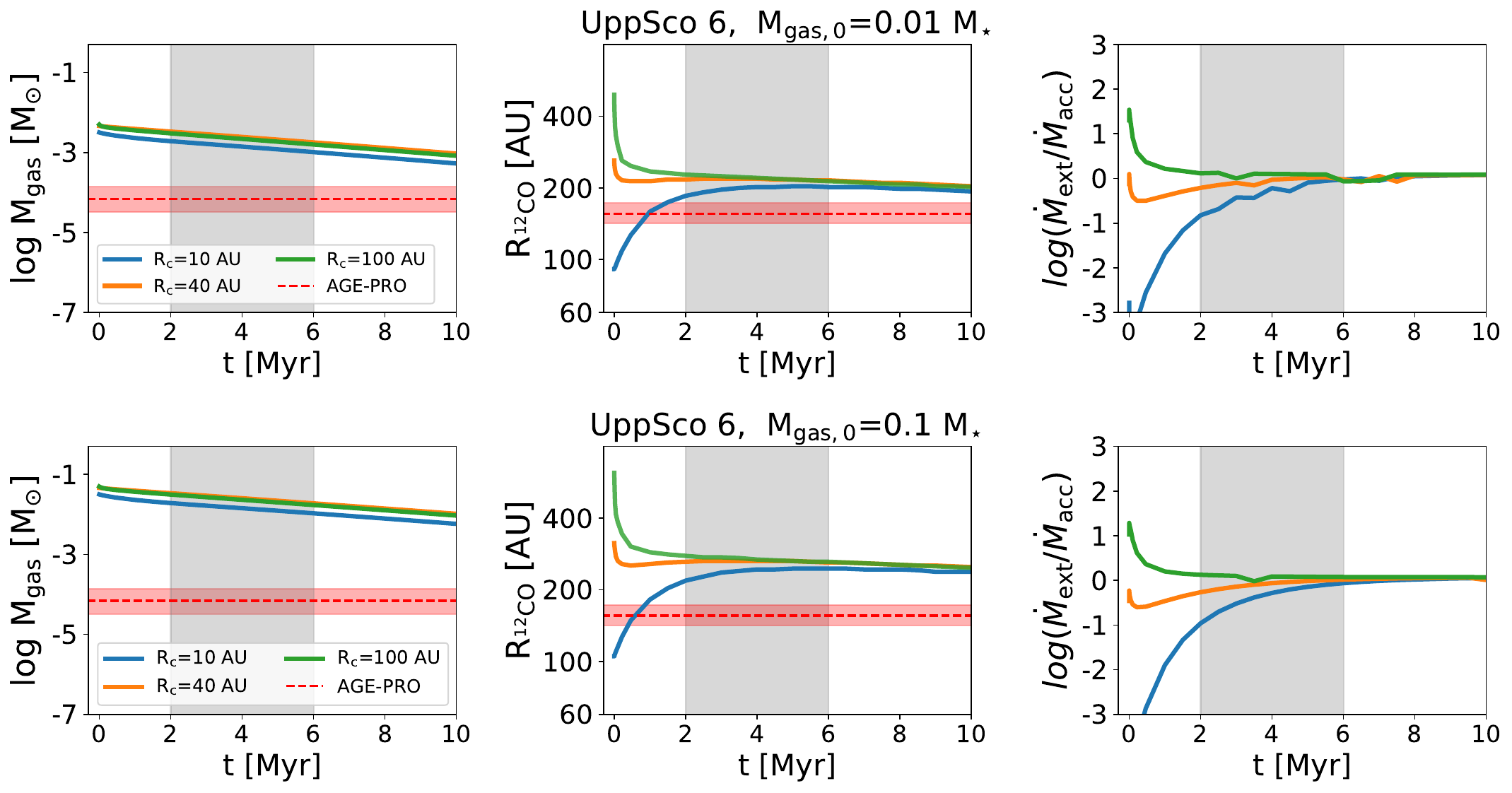}
        \caption{UppSco 6. Time evolution of the disk mass, radius, and  $\dot{M}_{\mathrm{ext}}/\dot{M}_{\mathrm{acc}}$, when the initial $R_{\mathrm{c}}$ is set to 10 AU, 40 AU, and 100 AU. The top and bottom rows refer to simulations in which the initial disk mass is $1\%$ and $10\%$ of the stellar mass, respectively. The AGE-PRO observations and uncertainties are shown in red.}
        \label{fig:us6}
    \end{figure*}
    \begin{figure*}
        \centering
        \includegraphics[width=1\linewidth]{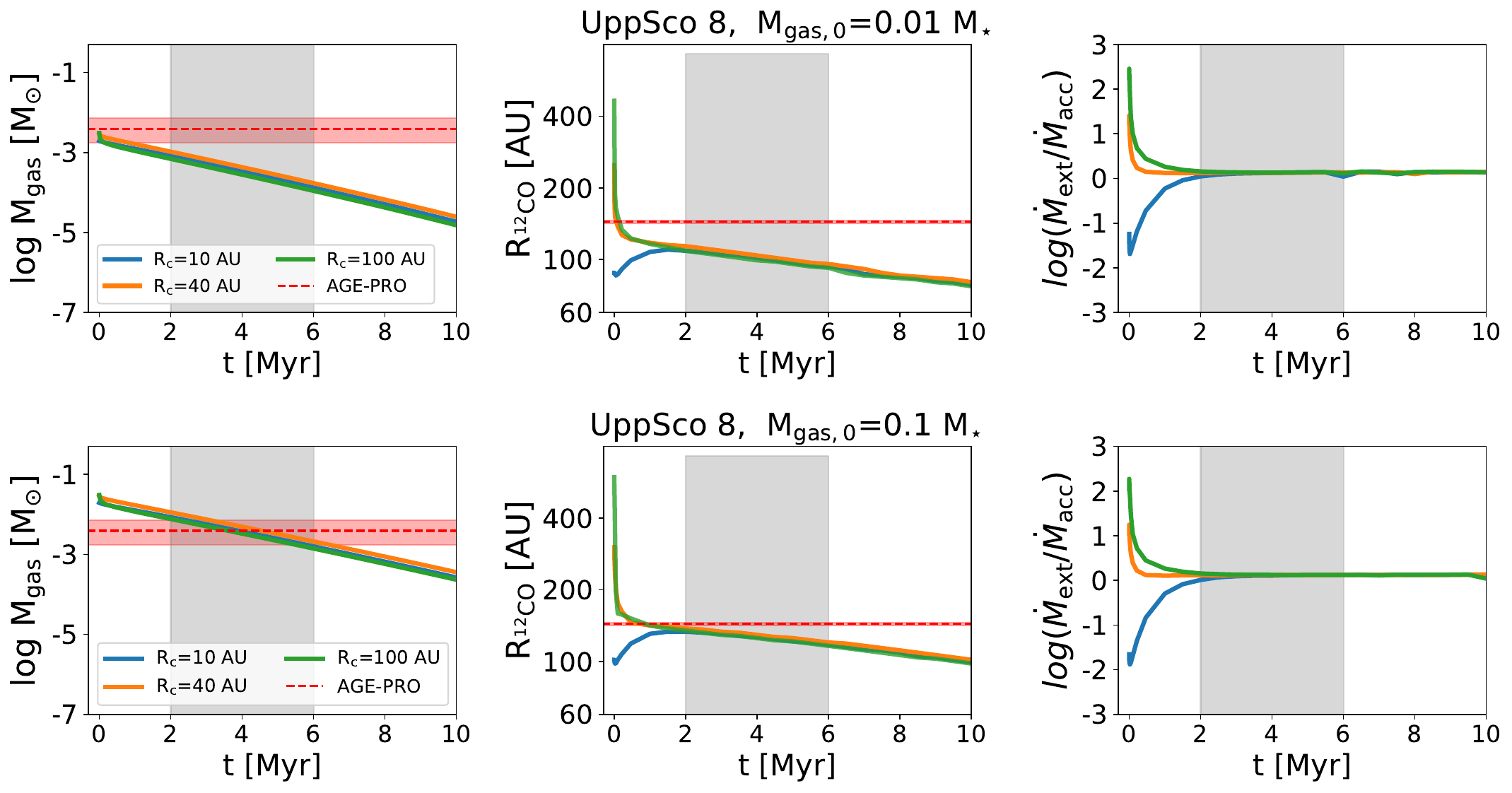}
        \caption{UppSco 8. Time evolution of the disk mass, radius, and  $\dot{M}_{\mathrm{ext}}/\dot{M}_{\mathrm{acc}}$, when the initial $R_{\mathrm{c}}$ is set to 10 AU, 40 AU, and 100 AU. The top and bottom rows refer to simulations in which the initial disk mass is $1\%$ and $10\%$ of the stellar mass, respectively. The AGE-PRO observations and uncertainties are shown in red.}
        \label{fig:us8}
    \end{figure*}
    We want to investigate how the initial disk extent influences the final value of disk mass and size, when an external FUV radiation field is introduced, and whether this parameter can be constrained by AGE-PRO observations. In this section we limit our analysis to just two sources from our sample, UppSco 6 and UppSco 8, which are among the disk with the lowest uncertainty on the evaluated FUV flux (see Figure \ref{fig:fuv_flux}), and allow the investigation of two different regimes of stellar mass and flux ( $\sim$0.50 $\rm{M}_{\odot}$ and FUV flux $\sim$4 $\rm{G}_{0}$ for UppSco 6, while $\sim$0.30 $\rm{M}_{\odot}$ and $\sim$10 $\rm{G}_{0}$ for UppSco 8.
    
    In Figure \ref{fig:us6} and \ref{fig:us8} we show the time evolution of disk mass, radius, and ratio between the mass loss rate due to external photoevaporation and the accretion rate onto the star ($\Dot{M}_{\rm{ext}} / \Dot{M}_{\rm{acc}}$), given different initial values for the characteristic radius $R_{\mathrm{c}}$.
    We found that disks with different initial extent evolve to a similar size in the estimated age of the region. On the other hand, our analysis reveals that the AGE-PRO gas disk masses are not precise enough to be used as potential constraint for the initial disk extent. The differences in final disk mass and final disk size for various initial disk extents, at older ages are too small to provide a robust constraint of the initial ${R}_{\rm{c}}$. In order to obtain reliable constraints on this parameter, the difference in radius and/or mass caused by varying the initial disk extent should be at least larger than the uncertainty on the observed values. This justifies our approach in which we set a fixed $R_{\rm{c},0}$ in our simulations.
    \\
    
    We verified that, after a few million years the disk achieves a state of equilibrium in which disk material is dispersing in winds at the same rate at which it is accreting onto the star. Although this behavior is well known from previous works (e.g. \citeauthor{Clarke_2007}, \citeyear{Clarke_2007}, \citeauthor{sellek_fried} \citeyear{sellek_fried}),  we demonstrate that disks with identical properties but distinct initial extents converge to a similar size, and then shrink, within the same timescale required for the disk to achieve the state of equilibrium.
    In particular, initially large disks are easily truncated by external photoevaporation (the weakly gravitational bound material at larger radii is easily removed, and the disk shrinks from the beginning of its lifetime). While, initially compact disks viscously spread until reaching the equilibrium between the photoevaporation rate and the mass accretion rate, then slowly decrease. 
    We provided a representative result based on only two sources; however, the same conclusions can be drawn for the entire sample.
    %
    %
    %
    %
\subsection{Correlations: FUV Flux and Disk Properties}\label{subsec:correlation}
\begin{figure*}[htb]
    \centering
    \begin{minipage}[b]{0.32\textwidth}
        \centering
        \includegraphics[width=\textwidth]{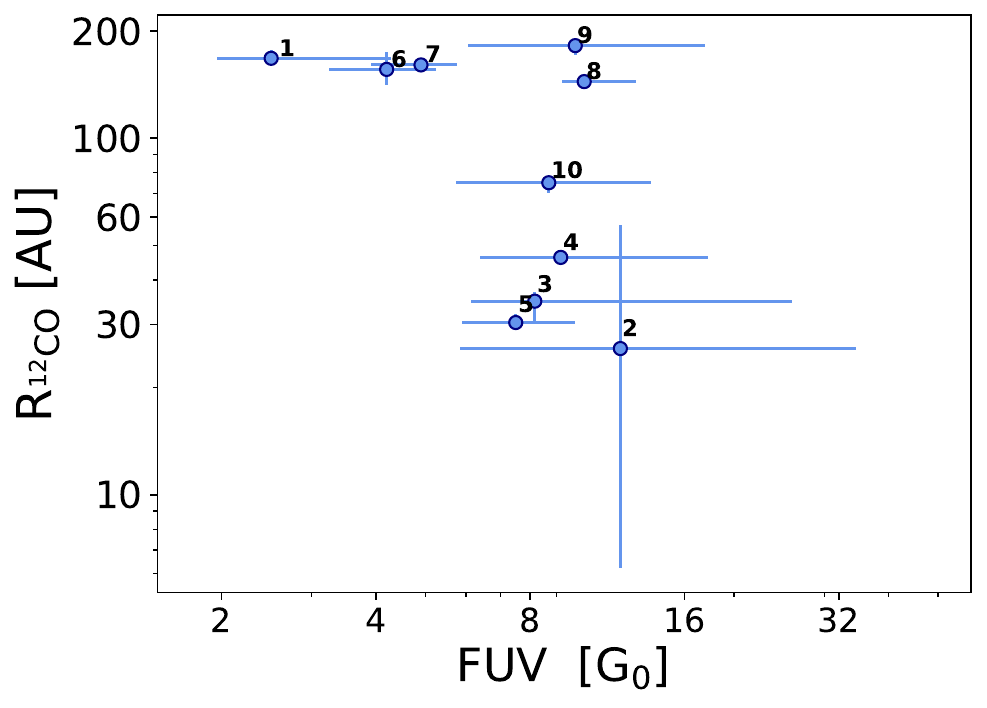}
    \end{minipage}
    \hspace{0.1\textwidth}
    \begin{minipage}[b]{0.32\textwidth}
        \centering
        \includegraphics[width=\textwidth]{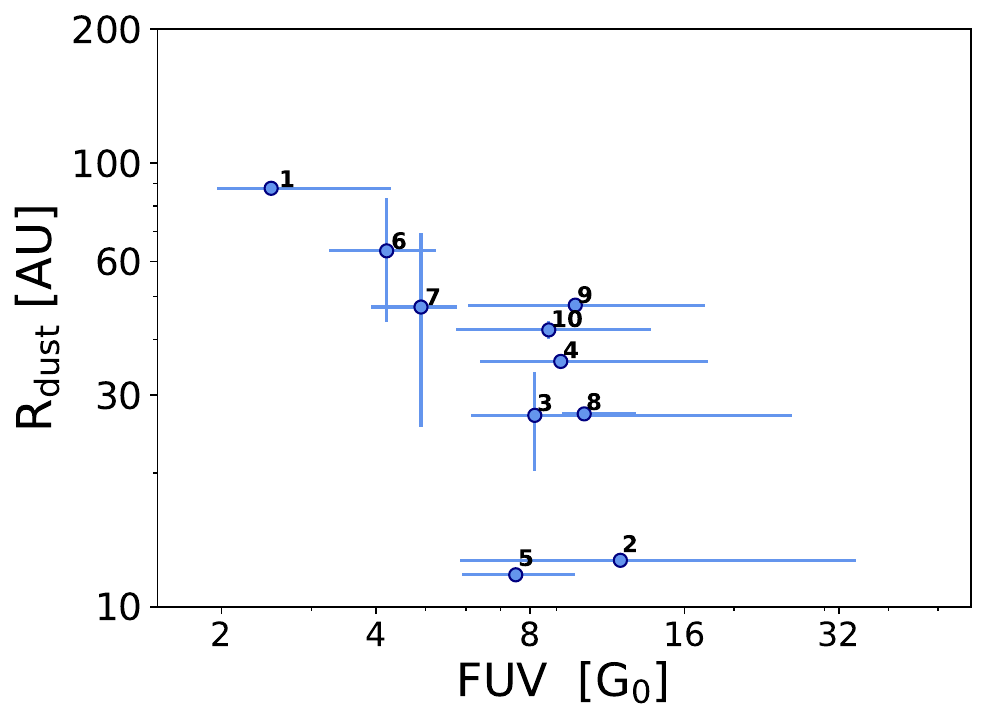}
    \end{minipage}
    \begin{minipage}[b]{0.32\textwidth}
        \centering
        \includegraphics[width=\textwidth]{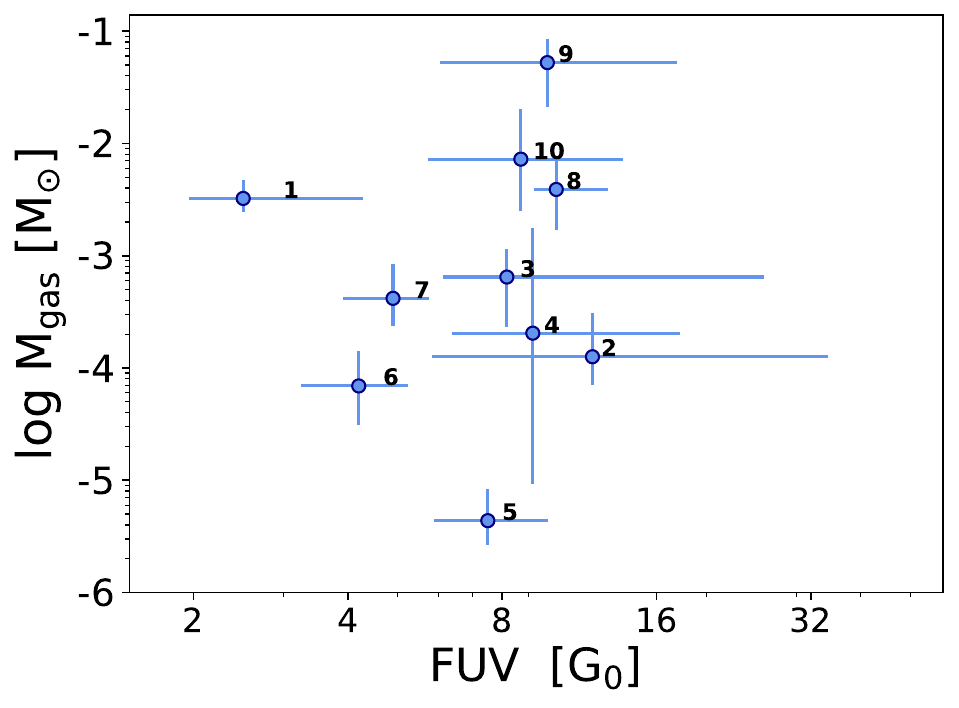}
    \end{minipage}
    \hspace{0.1\textwidth}
    \begin{minipage}[b]{0.31\textwidth}
        \centering
        \includegraphics[width=\textwidth]{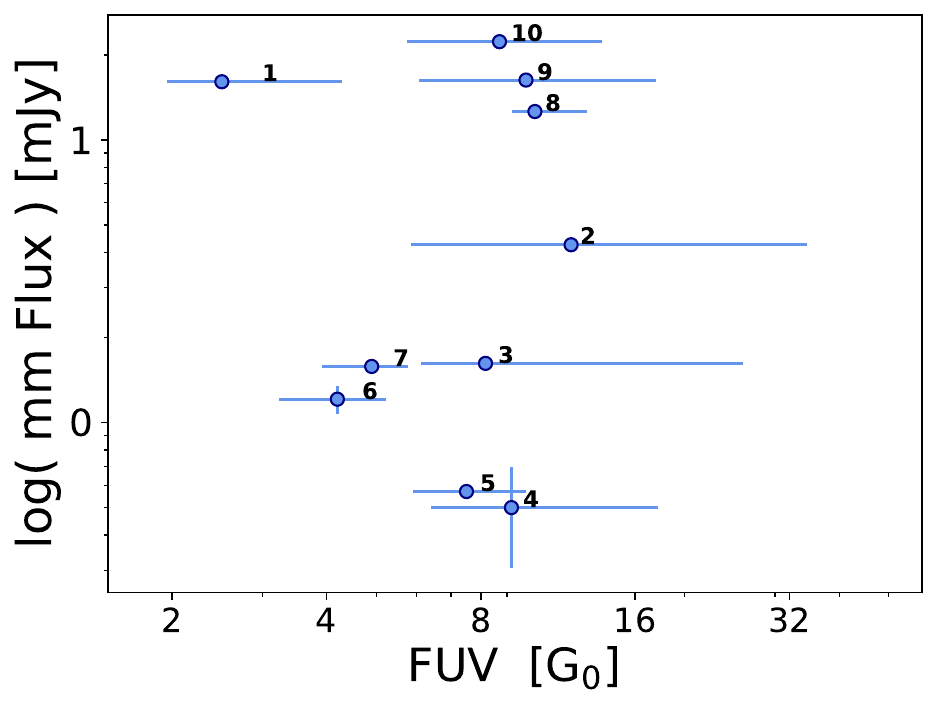}
    \end{minipage}
    \caption{The estimated FUV flux experienced   by the ten AGE-PRO sources in Upper Sco is plotted against the gas radius (\textit{top left}), dust radius (\textit{top right}), gas mass (\textit{bottom left}), 1.3 mm flux (\textit{bottom right}). The disk properties are those retrieved from the AGE-PRO observations. The correlation between size and FUV flux is marginal, while no correlation is found with the mass.}
    \label{fig:corr_Rgas_Rdust_FUV}
\end{figure*}
In this Section we search for potential correlations between the FUV flux experienced by the ten targets in the Upper Sco region, and the disk masses and radii observed by AGE-PRO.
In order to do this, we use the Spearman's coefficients, where the $\rho$ parameter refers to steepness of the correlation, which can be positive or negative, and the $p$-value indicates the strength of the correlation, being considered strong or weak depending on whether it is smaller or greater than 0.05, respectively. 
The Spearman's parameters, resulting from applying a Monte Carlo method, and the corresponding 16$^{\mathrm{th}}$ and 84$^{\mathrm{th}}$ percentile of the distributions, are listed in Table \ref{tab:correlation_parameters}.
We find a weak negative correlation between the disk sizes, both gas and dust, and FUV fluxes, whereas disk masses show no apparent correlation with the external radiation (where $p$-values are not smaller than 0.05), as can be seen also from the values plotted in Figure \ref{fig:corr_Rgas_Rdust_FUV}. However, our disk sample is too small to offer robust statistical evidence for the presence of correlations. Therefore, this analysis will be extended in future with a larger sample of disks.

    \begin{deluxetable}{ccccccc}
        \tablecaption{Spearman's correlation parameters \label{tab:correlation_parameters}}
        \tablehead{ \colhead{ Quantity } & \colhead{ $\langle \rho \rangle$} & \colhead{$\rho$,16\%} & \colhead{$\rho$,84\%} & \colhead{ $\langle p \rangle$} & \colhead{p,16\%} & \colhead{p,84\%}
        }
        \startdata
        $R_{^{12}\rm{CO}}$ & -0.41  & -0.63 & -0.18 & 0.12 & 0.02 & 0.31 \\ 
        $R_{\rm{dust}}$ & -0.52 & -0.69 & -0.28 & 0.064 & 0.01 & 0.21\\ 
        $M_{\rm{gas}}$ & -0.07 & -0.28 & 0.12  & 0.53 & 0.25  & 0.61  \\
        $M_{\rm{dust}}$ & -0.12 & -0.40 & 0.05  & 0.41 & 0.22 & 0.55  \\ 
        \enddata
        \tablecomments{Median, 16$^{\mathrm{th}}$ and 84$^{\mathrm{th}}$ percentiles related to the $\rho$ and $p$ Spearman's values for testing a potential correlation between FUV flux and disk radius (gas and dust) and mass (gas and dust).}
    \end{deluxetable}
\section{Discussion}\label{sec:discussion}

\subsection{Gas Results: the Importance of Considering External Photoevaporation}
The main outcome arising from the comparison between models and observations is that the sole reliance on a purely viscous model is insufficient to account for the observed disk masses and radii in Upper Sco. 
This result agrees with the findings from population synthesis investigation conducted by \citet{AGEPRO_population_benoit}.  
The process of external photoevaporation reveals to be relevant in this region, influencing mostly disk radii over disk masses, even if only a moderate level of FUV radiation is present. 
Disks in lower and higher irradiated star forming regions have different evolutionary history, and therefore must be modelled separately. Considering that the average FUV flux experienced by disks in the Upper Sco region is higher than in the Lupus region, the average properties of Upper Sco disks cannot be regarded as representative of old evolutionary stages of Lupus disks, as pointed out also in the work of \citet{Zagaria_2023_UpperSco}.
\\

From the comparison between models and observations we note a general preference for initially low-mass disks. A more detailed analysis can be conducted as follows. From the observed gas masses and the outcomes of the model (Figure \ref{fig:mass_gas} and \ref{fig:diff_mass_gas}), we deduce that six targets out of ten show a preference for initially low-mass disks (closer to 0.01 $\rm{M}_{\star}$ than 0.1 $\rm{M}_{\star}$). These are UppSco 1, UppSco 3, UppSco 4, UppSco 5, UppSco 6, and UppSco 7. UppSco 2, UppSco 8 UppSco 9 and UppSco 10 are better described instead by initially more massive disks (closer to 0.1 $\rm{M}_{\star}$). 
Models overestimate the observed disk mass and size of UppSco 5, when external photoevaporation is included. Gas size is overestimated for UppSco 3 and UppSco 4 as well, with the latest showing gas mass consistent with the model within the error bars (see top panel in Figure \ref{fig:mass_gas}).
The comparison of gas sizes, shown in Figure \ref{fig:rco}, is crucial to investigate the impact of external photoevaporation. However, sizes do not serve as a reliable discriminant for the initial disk mass. The difference between the two considered evolutionary scenarios is indeed minimal.

Considering the current debate regarding the main driving mechanism of disk accretion (viscous vs MHD winds), the failure of the viscous model to explain the small disk sizes observed in Upper Sco should not be interpreted as a demonstration that accretion in this region is driven by MHD winds, although winds may be common (see \citeauthor{Fang_pascucci_2023} \citeyear{Fang_pascucci_2023}, \citeauthor{AGEPRO_population_benoit} \citeyear{AGEPRO_population_benoit}). Rather, it underlines the importance of considering external effects when describing disk evolution, and adds another parameter (the irradiation level) to the comparison of disk properties across different SFRs, which are typically compared on the basis of the age alone. On the other hand, the relative success of the viscous external photoevaporation model in reproducing disk sizes in Upper Sco should not be interpreted as a proof that disk accretion is driven by viscosity, since we did not test alternative models in which accretion is driven by winds. Furthermore, we are aware of the results previously shown by \citet{Trapman_mhd}, and more recently by \citet{AGEPRO_population_benoit}, which highlight that a pure MHD wind-driven model faces challenges in reproducing the observed disk sizes in this particular region, if we refer to population synthesis models set to match disk properties at younger ages as seen in the Lupus region.
\subsection{Dust Results: the Necessity of Including Unresolved Substructures}
The results from both mm fluxes and dust radii suggest a higher preference for initially more massive disks (see Figure \ref{fig:mm_flux}) compared to what deduced from the gas-only simulations. This result is likely caused by the absence of substructures in our prescription. Substructures are known to impede the radial drift of dust, thereby preventing the expected decrease in dust mass and size. In this work we consider smooth disks and, from the work of \citet{sellek_fried} we know that dust drifts even more efficiently in disks subject to external photoevaporation and \citet{AGEPRO_dust_evolution_Nico} found that weak or strong substructures are needed to explain AGE-PRO observations. From the comparison between gas and dust sizes (see Figure \ref{fig:rco_rdust}), we conclude that, even if disks are truncated in gas due to external photoevaporation, substructures are still needed to explain the observed gas-to-dust size ratios.
Planetesimals formation, gravitational instabilities (\citeauthor{dust_dynamics_2012}, \citeyear{dust_dynamics_2010}, \citeauthor{Pinilla_2012},\citeyear{Pinilla_2012}), and planet formation (which is reasonable to be occurred for, at least, some of our targets \citeauthor{Sierra_usco1} \citeyear{Sierra_usco1}), can cause disk pressure bumps trapping dust grains and delaying radial drift.
In presence of an external FUV radiation field, the study of \citet{Garate_2023} suggests that dust traps only survive photoevaporation if they are located inside the truncation radius. Otherwise these disperse along with the gas even in cases where the experienced radiation increases from milder to stronger values. 

Among the ten disks in our sample, UppSco 1 and UppSco 7 (see \citeauthor{AGEPRO_Upp7} \citeyear{AGEPRO_Upp7}) exhibit clear substructures in the continuum at the current resolution, while a detailed study in the visibilities conducted in the Fourier space by \citet{AGEPRO_X_dust_disks}, reveals the presence of substructures in UppSco 6, UppSco 8, UppSco 9 and UppSco 10 as well. Given that the spatial resolution of AGE-PRO is $\sim$30 AU \citep{AGEPRO_IV_UpperSco}, we expect all (or most) of our targets to show substructures at smaller spatial scales than those that could be accessed by our observations. 
Future high-resolution observations will be able to confirm this hypothesis.
Moreover, in \citet{AGEPRO_X_dust_disks} is shown that no evolution of the dust disk radius with age is detected, while we observe a potential correlation with the FUV flux (Section \ref{subsec:correlation}), which increases the strength of our argument that external photoevaporation cannot be neglected in Upper Sco.
\\

We want to highlight that, while an underestimate of the observed dust radius can be explained with the fact that we did not include substructures able to stop the dust drift, an overestimate of dust disk size cannot be explained without changing global parameters (e.g. the alpha viscosity parameter, though this presents the additional problem of how to explain the observed disk masses, as described in \ref{subsec:alpha}). 
Dust results are particularly useful in constraining the initial disk mass in the case of UppSco 1. This is the most extended disk in the continuum among the AGE-PRO targets in Upper Sco, showing a bright ring-like substructure \citep{Sierra_usco1}. Since our models do not include substructures, we expect the actual dust size to be underestimated. Consequently, from the plots in Figure \ref{fig:rdust}, this target is unlikely to have a high initial disk mass. 

\subsection{Additional Factors Impacting Disk Masses and Sizes}\label{subsec: discussion_additional_factors}
Several processes, including variations in the FUV flux experienced by the disk, and the loss of material in photoevaporative winds, can impact the resulting disk masses and sizes.
In this Section we briefly discuss some of the main caveats and their consequences on disk evolution.

    \textbf{Time-varying FUV radiation field.} \quad   The FUV flux experienced by a disk during its lifetime is not constant in time, especially due to the evolution in FUV luminosity of the massive stars and the dynamics of the cluster.
    
\begin{figure}[htbp]
    \centering
    \begin{minipage}[b]{0.45\textwidth}
        \centering
        \includegraphics[width=\textwidth]{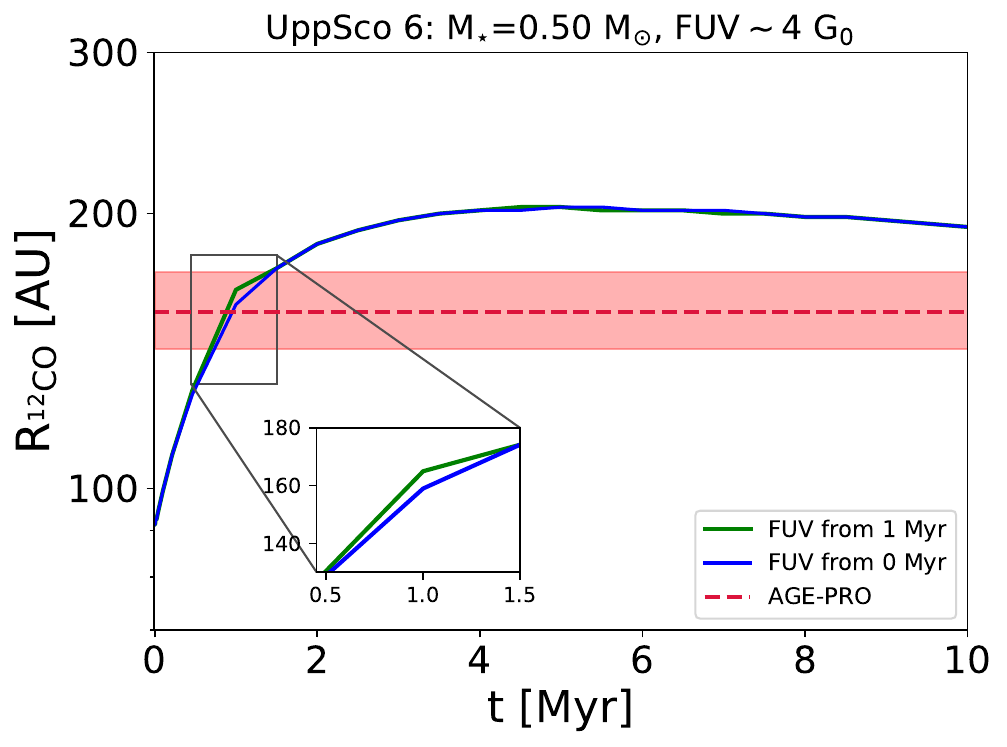}
    \end{minipage}
    \hfill
    \hspace{0.002\textwidth}
    \begin{minipage}[b]{0.45\textwidth}
        \centering
        \includegraphics[width=\textwidth]{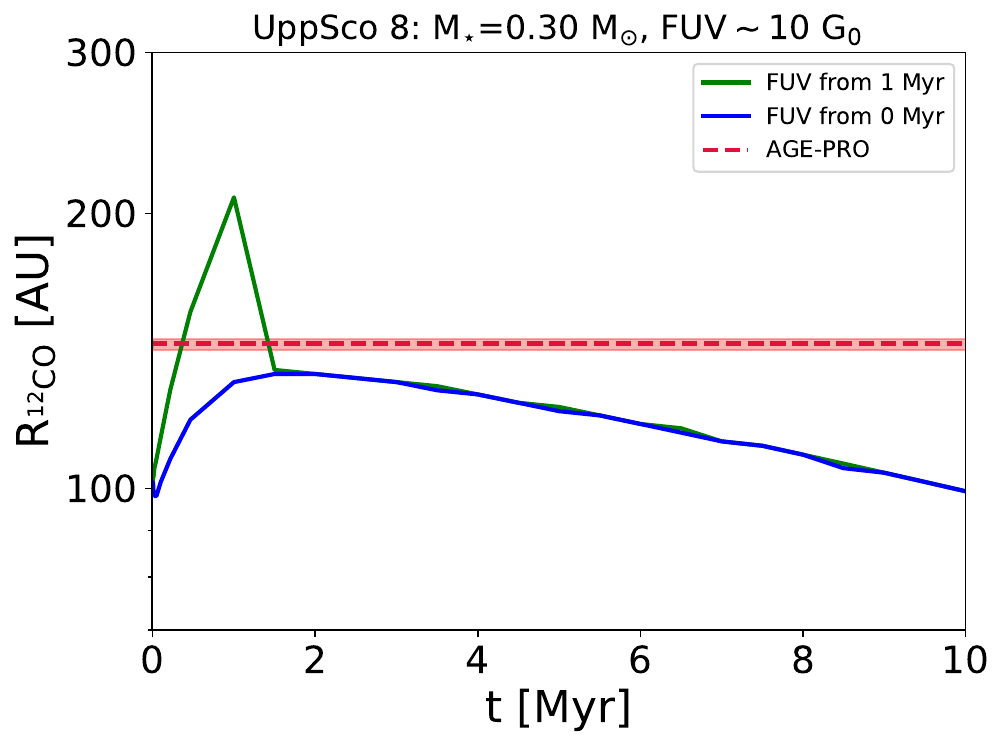}
    \end{minipage}
    \caption{Time evolution of the gas radius when a constant FUV flux is experienced from the beginning of the disk lifetime, and after 1 Myr (shown in blue and green lines, respectively). Application to the case of UppSco 6 (\textit{top panel}) and UppSco 8 (\textit{bottom panel}), with initial parameters: R$_{\rm{c,0}}$=10 AU, $\alpha = 10^{-3}$, and M$_{disk,0} = 0.01 \rm{M}_{\star}$ for UppSco 6, and M$_{disk,0} = 0.1 \rm{M}_{\star}$ for UppSco 8.}
    \label{fig:late_rco}
    \end{figure}

    The work of \citet{Kunitomo_2021} shows that intermediate-mass stars, such as those found around Upper Sco, exhibit a sharp increase in FUV luminosity after a period of time that depends on the stellar mass, and then remains approximately constant. Given that the average mass of the B stars in Upper Sco is $\sim$3$ \ \rm{M}_{\odot}$, from the stellar evolution models of \citet{Kunitomo_2021} based on the effective temperature of the stars (see Figure 3 in \citeauthor{Kunitomo_2021} \citeyear{Kunitomo_2021}), we expect the stellar FUV luminosity to increase from $\sim$$10^{30} \ \rm{erg} \ \rm{s}^{-1}$ to $\sim$$10^{34} \ \rm{erg} \ \rm{s}^{-1}$ after the first million year.    
    This means that, at a distance of 0.2 pc, the FUV flux experienced by the disk will shift from $\sim$$10^{-4} \ \rm{G}_{0}$ to $\sim$$7 \ \rm{G}_{0}$. 
    In addition, during (approximately) the first million year of its lifetime the disk is well shielded against external UV radiation by interstellar gas (\citeauthor{Qiao_extinction_2022} \citeyear{Qiao_extinction_2022}, \citeauthor{Ali_2019} \citeyear{Ali_2019}). In order to evaluate the effects of this delayed exposure to an external radiation, we perform a test simulation in which the external FUV flux is turned on after 1 Myr.
    The resulting evolution of the disk radius is then compared to that of a simulation in which the FUV radiation field is present from the beginning. 
    We focus on the comparison of disk radii. 
    As a representative example, we perform this test on UppSco 6 and UppSco 8. The results are shown in Figure \ref{fig:late_rco}.  
    We find that when external photoevaporation is activated the radius rapidly evolves to the same value predicted if the disk experiences the same external radiation from the beginning. Therefore, while this process may be influential in younger star-forming regions, in the context of disks in Upper Sco, they are not affected by the timing of the actual exposure to the estimated FUV flux.

    The external radiation experienced by disks can fluctuate over time due to the stellar cluster dynamics (e.g. \citeauthor{Wilhelm_shielding} \citeyear{Wilhelm_shielding}), as the distance between objects in the same cluster does not remain fixed in time.
    As the cluster expands, stars are exposed to a weaker external radiation with time. Furthermore, it is reasonable to infer that the average FUV radiation field was stronger in the past, given the presence of O-type stars (now turned into neutron stars).
    Nevertheless, 
    disks detected in a high irradiated environment, which are expected to be smaller and less massive or even dispersed, could have migrated from a low irradiated region \citep{winter_2019b}, and vice-versa. This last scenario may be involved in the explanation for the detectability of UppSco 2, which is predicted to by fully dispersed in less than 6 Myr.
    On the other hand, weakly irradiated disks might have been formed near O-type stars, experiencing high irradiation, and then drifted away over time. Future investigations of the 3D geometry and kinematics of the region could address this topic in more detail.
    
    \textbf{Winds.} 
    High-energy radiation from the central star (mostly EUV and X-ray) and magnetized winds launched from the disk (MHD) may contribute significantly to the disk dispersal.
    
    EUV photons from the host star drive late-stage disk evolution, causing gap formation when accretion matches photoevaporative mass loss, leading to rapid disk dispersal in $\sim 10^{5}$ yr \citep{Clarke_euv}.
    \citet{picogna_mass_dependence} showed that the X-ray irradiation from the central star produces a mass-loss rate which scales linearly with the stellar mass. Our AGE-PRO sample covers stellar masses between $\sim$$0.2 \ {M}_{\odot}$ and $\sim$$0.6 \ \rm{M}_{\odot}$, corresponding to a luminosity $\rm{L}_{\rm{X}} \sim 10^{29} \ \rm{erg} \ \rm{s}^{-1}$ and mass loss rates between $\sim$$5 \times 10^{-9}$ and $\sim$$10^{-8} \ \rm{M}_{\odot} \rm{yr}^{-1}$. Our models do not include the inside-out depletion of material due to internal photoevaporation, which may impact disk mass, accretion rate, and cavity formation.
    
    In three or our AGE-PRO disks (UppSco 1, UppSco 3 and UppSco 10), \citet{Fang_pascucci_2023} detected emission of the [OI] $\lambda$ 6300 line, tracing the presence of MHD-driven winds which can lead to an additional decrease of disk masses and sizes. The impact of this process on the dynamics of the analyzed sources is still to be quantified.

\section{Conclusion}\label{sec:conclusion}
In this work we showed that external photoevaporation can significantly impact the evolution of moderately irradiated protoplanetary disks, experiencing FUV fluxes greater than 1 $\mathrm{G}_{0}$. We compared the observed disk masses and sizes of the ten AGE-PRO targets in Upper Sco, with those predicted by models that incorporate both viscous evolution and external photoevaporation. 
The actual FUV flux experienced by each disk was computed accounting for the contribution of all the OB massive stars in the region. 
We performed numerical simulations varying the initial disk mass, the initial disk extent, and the viscosity. Then, we compared disk masses and radii at 2 and 6 Myr (i.e. average range of ages for the selected Upper Sco targets), with the AGE-PRO observations. In particular, we focused on $^{12}$CO emission and millimeter flux to investigate respectively the gas and dust component. 
The main outcomes of this work are summarized as follows:
\begin{enumerate}
    \item The AGE-PRO sources in Upper Sco experience median FUV fluxes between 2 and 12 $\rm{G}_{0}$.
    \item Protoplanetary disks are truncated by external photoevaporation, and consequently shrink, even at moderate levels of irradiation (1 - 10 G$_{0}$). The process of external photoevaporation is not negligible in Upper Sco, where we demonstrated a substantial decrease in gas disk sizes. As a consequence, the pure viscous framework is unable to reproduce the observed disk properties in this region. 
    \item The relative success in reproducing gas sizes through the inclusion of external photoevaporation to our model underlines that different SFRs cannot be compared on the sole basis of the age. Indeed, we showed that the irradiation level is a crucial parameter even in a moderate irradiated environment. In other words, the average disk properties of the Upper Sco region cannot be treated as an old version of the Ophiuchus and Lupus regions.
    \item The initial disk extent cannot be constrained by comparing models that include external photoevaporation and observations. Disks subject to the same FUV flux and with different initial extent converge to a similar size by the age of Upper Sco. On a similar timescale disks reach the condition in which the mass loss rate due to external photoevaporation equals the accretion rate onto the star. 
    \item Dust masses and radii are in general not well reproduced by our models, indicating the presence (and the relevance) of unconsidered processes that are influencing the radial drift, such as dust traps and substructures. Even if the inclusion of photoevaporation results in smaller disks in gas, substructures are still needed to explain the observed gas-to-dust size ratios.   
    \item Models with the inclusion of external photoevaporation provide a good representation of the actual gas size observations (within a factor $<$2) in 7 out of 10 disks (our model predicts larger radii for UppSco 3, UppSco 4 and UppSco 5), considering initial disk masses between 1\% and 10\% of the stellar mass. 
    In general, initially low-mass disks (0.01 $\rm{M}_{\star}$) are preferred.    
\end{enumerate}

\section*{Acknowledgments}
We thank the referee for the careful reading of the manuscript and for providing useful comments and suggestions that helped us improve our understanding of the science case. We thank Dr. Andrew Sellek for the discussion and the suggestions regarding the numerical implementation of the process of external photoevaporation in the numerical code. R.A. and G.R. acknowledge funding from the Fondazione Cariplo, grant No. 2022-1217, and the European Research Council (ERC) under the European Union’s Horizon Europe Research and Innovation Programme under grant agreement No. 101039651 (DiscEvol). Views and opinions expressed are, however, those of the author(s) only and do not necessarily reflect those of the European Union or the European Research Council Executive Agency. Neither the European Union nor the granting authority can be held responsible for them. 
P.P. acknowledges the support from the UK Research and Innovation (UKRI) under the UK government’s Horizon Europe funding guarantee from ERC (under grant agreement No. 101076489). L.T. acknowledges the support of NSF AAG grant No. 2205617. K.Z. acknowledges the support of NSF AAG grant No. 2205617. I.P. and D.D. acknowledge support from the Collaborative NSF Astronomy and Astrophysics Research grant (ID: 2205870). 
L.A.C. acknowledges support from the Millennium Nucleus on Young Exoplanets and their Moons (YEMS), ANID-Center Code NCN2024$\_$001, and FONDECYT grant No. 1241056. 
A.S. acknowledges support from FONDECYT de Postdoctorado 2022 No. 3220495 and support from UKRI under the UK government's Horizon Europe funding guarantee from ERC (under grant agreement No 101076489). J.M. acknowledges support from FONDECYT de Postdoctorado 2024 No. 3240612. 
L.P. acknowledges support from ANID BASAL project FB210003 and ANID FONDECYT Regular No. 1221442. B.T. acknowledges support from the Programme National
$"$Physique et Chimie du Milieu Interstellaire$"$ (PCMI) of CNRS/INSU with INC/INP and co-funded by CNES. 
C.A.G. acknowledges support from FONDECYT de Postdoctorado 2021 No. 3210520.
C.G-R. acknowledges support from YEMS, ANID-Center Code NCN2024$\_$001. 
This paper makes use of the following ALMA data: ADS/JAO.ALMA$\#$2021.1.00128.L. ALMA is a partnership of ESO (representing its member states), NSF (USA) and NINS (Japan), together with NRC (Canada), MOST and ASIAA (Taiwan), and KASI (Republic of Korea), in cooperation with the Republic of Chile. The Joint ALMA Observatory is operated by ESO, AUI/NRAO and NAOJ. The National Radio Astronomy Observatory is a facility of the National Science Foundation operated under cooperative agreement by Associated Universities, Inc.

\vspace{5mm}
\software{Astropy (\citeauthor{astropy:2013} \citeyear{astropy:2013}, \citeyear{astropy:2018}), \texttt{Dustpy} \citep{dustpy}, Matplotlib \citep{Matplotlib_2007}, Numpy \citep{Numpy_2020}}

\clearpage

\appendix
\section{OB stars and FUV flux distribution and uncertainties}
In this Appendix, we primarily provide additional information on the sample of massive stars used to compute the FUV flux at each star-hosting disk position. 
Then, we show the full 1D probability distribution of FUV fluxes from which we retrieved median and uncertainties summarised in Table \ref{tab:mstar_fuv} and shown in Fig. \ref{fig:fuv_flux}.
Finally, we discuss the uncertainties in the FUV flux calculation.

\subsection{OB stars}\label{appendix:ob_stars}
The OB stars responsible for the FUV flux experienced by the AGE-PRO sources in Upper Sco are listed in Table \ref{tab:OB_stars_pars}, along with their RA and Dec positions, parallax, parallax error, spectral type, and FUV luminosity.
\begin{table}[ht]
    \caption{Parameters of the OB-type stars in a 3D box around Upper Sco}
    \label{tab:OB_stars_pars}
    \def\arraystretch{1.2}
    \begin{tabular*}{0.97\columnwidth}{cccccccc}
    \hline
    \hline
    SIMBAD id & Gaia DR3 id & RA [deg] & Dec [deg] & plx [mas] & plx err [mas] & SpT & $L_{\mathrm{FUV}}$ [erg s$^{-1}$] \\ \hline
    *   2 Sco & - & 238.403 & -25.3271 & 6.49 & 0.51 & B2.5Vn & 3.467e+36 \\ 
    *   3 Sco & 6235747125966268928 & 238.6647 & -25.2437 & 6.9332 & 0.0542 & B8III/IV & 9.852e+34 \\ 
    * rho Sco & 6041076668127513984 & 239.2212 & -29.2141 & 6.91 & 0.19 & B2IV-V & 3.467e+36 \\ 
    * sig Sco B & 6048602103662751488 & 245.291 & -25.5926 & 7.2334 & 0.1772 & B9.5V & 2.451e+34 \\
    V* V933 Sco & 6244725050721030528 & 245.0229 & -20.0564 & 7.6496 & 0.0297 & B9II/III & 2.451e+34 \\ 
    * pi. Sco & 6235406071207202688 & 239.713 & -26.1141 & 5.57 & 0.64 & B1V& 1.966e+37 \\
    * del Lup & 6005678131793146752 & 230.343 & -40.6475 & 6.7296 & 0.4819 & B1.5IV & 1.966e+37 \\ 
    * del Sco & - & 240.0834 & -22.6217 & 6.64 & 0.89 & B0.3IV & 6.968e+37 \\
    ... & ... & ... & ... & ...& ... & ... & ... \\
    \hline
    \end{tabular*}
    \begin{minipage}{0.87\columnwidth}
    \vspace{0.1cm}{\footnotesize{\textbf{Notes:} From left to right we provide SIMBAD and Gaia DR3 identification name of the OB-type stars, RA and Dec position, parallax and parallax error as provided by Gaia DR3 or Hipparcos where Gaia data are not available, spectral type, and FUV luminosity computed as described in Sec. \ref{sec: FUV calculation}. The full table is available as supplementary material.}}
    \end{minipage}
\end{table}
\subsection{FUV flux distribution}\label{appendix:flux_distr}
\begin{figure}
    \centering
    \includegraphics[width= 1\linewidth]{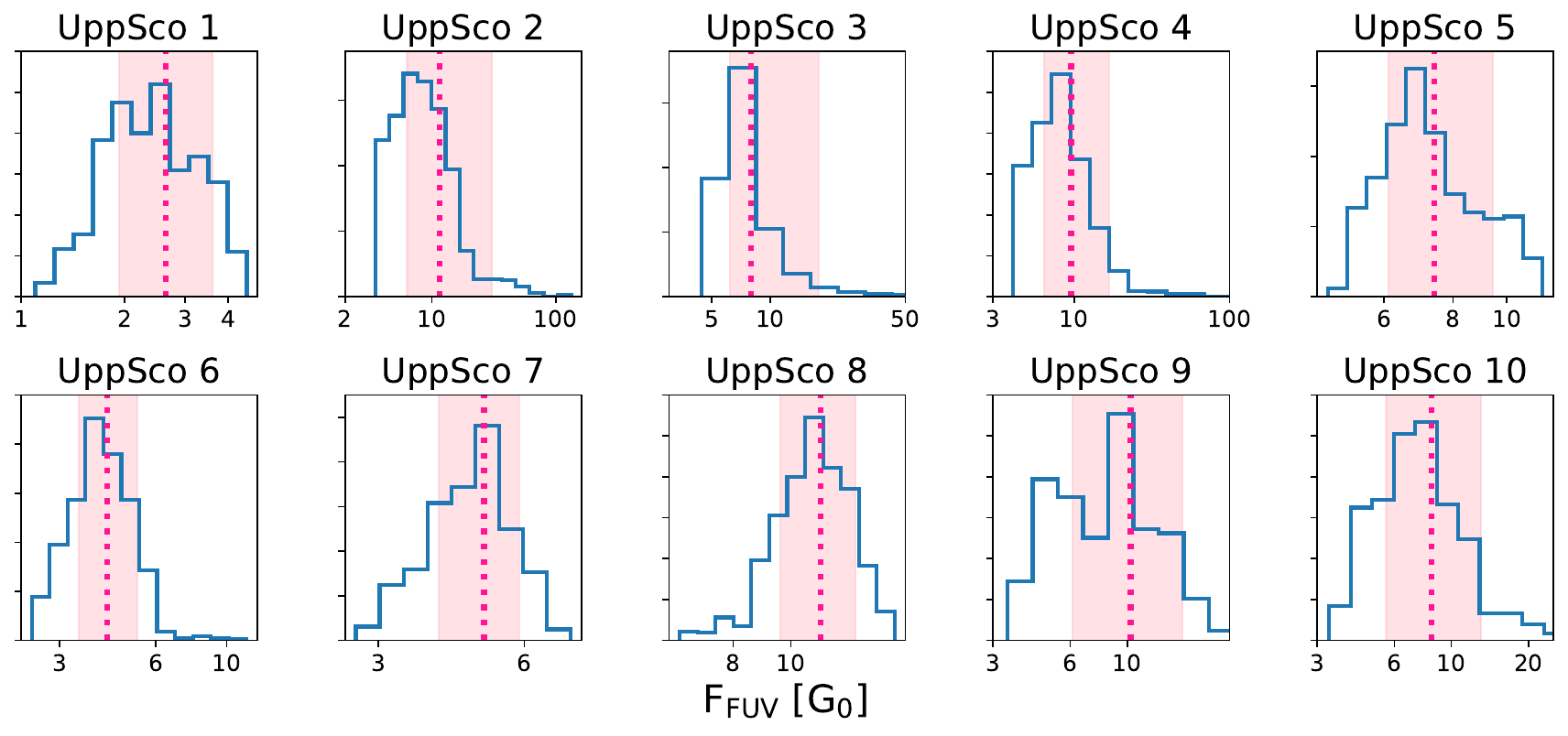}
    \caption{For each AGE-PRO target in Upper Sco, we show the distribution of FUV fluxes as resulting from applying the method detailed in Sec. \ref{sec: FUV calculation}.  }
    \label{fig:full_distributions}
\end{figure}

The FUV flux is computed using a Monte Carlo approach, sampling from parallax  full distributions of computed FUV fluxes.

\subsection{Uncertainty on the FUV flux calculation }\label{Appendix:flux_error}
    The error bars in the provided FUV flux estimates result from the uncertainties on the parallax measurements, which we use to compute the distance between the T Tauri and massive stars.    
    Despite the precision of the Gaia and Hipparcos data, large uncertainties on the parallax can arise when dealing with very bright stars (O and B stars) and/or faint stars (faint T Tauri stars).
    When the uncertainties on the line-of-sight distances of the objects are comparable to the physical separations between the OB stars and disks, which can occur when a massive star is close to a disk, the inferred median FUV flux may tend to be underestimated.
    A good example is provided by UppSco 2, which is $\sim0.37$ deg away from the B1 star \emph{$\omega$Sco}. Using a Monte Carlo sampling of parallax distributions, the median relative distance (in three-dimensional space) is estimated to be $\sim$6 pc, with 16\% and 84\% of the distribution at $\sim$2 pc and $\sim$12 pc, respectively. This yields the FUV flux to fluctuate between $\sim$5.8 G$_{0}$ and $\sim$31 G$_{0}$, with median of the distribution peaked at $\sim$12 G$_{0}$. Remarkably, \emph{$\omega$Sco} contributes to the $\sim 99 \%$ of the total FUV flux experienced by UppSco 2.    
    
\begin{figure}[htbp]
    \centering
    \includegraphics[width=0.35\textwidth]{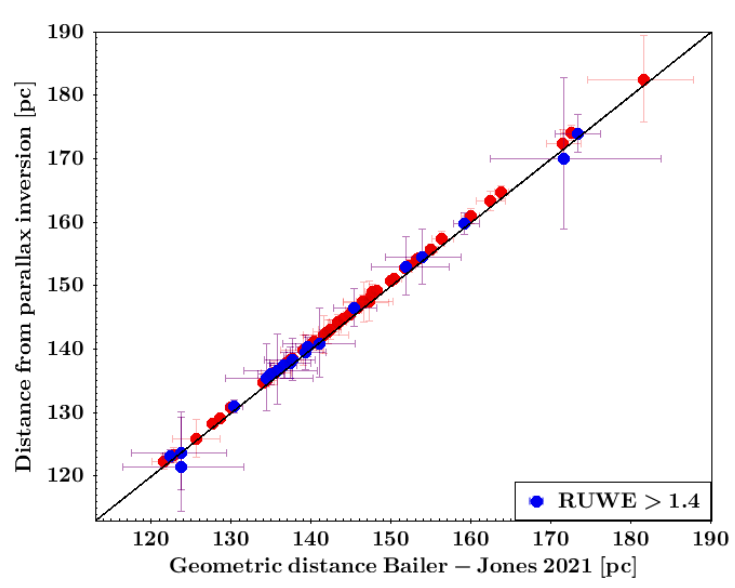}
    \caption{Comparison between distances obtained with the parallax inversion and the Bailer-Jones method. Blue dots mark the sources showing RUWE$>1.4$ (i.e. with not good parallax measurements), while red dots refer to sources with good defined parallax measurements.}
    \label{fig:ruwe}
    \end{figure}

    Previous works state that reliable line-of-sight distances cannot be obtained by a rough inversion of the parallax (e.g. \citeauthor{distance_from_kinematics} \citeyear{distance_from_kinematics}; \citeauthor{Bailer-Jones} \citeyear{Bailer-Jones}). However, Upper Sco is close enough to us to ensure good Gaia and Hipparcos parallax measurements, and we show in Figure \ref{fig:ruwe} that the ours and Bailer-Jones \citep{Bailer-Jones} method for distance evaluation yield the same outcome, confirming the validity of our approach. 
    We tested the astrometric goodness-of-fit associated with the parallax measurements considering the Renormalised Unit Weight Error (RUWE). As shown in Figure \ref{fig:ruwe}, the vast majority of the stars in our sample exhibit RUWE $< 1.4$, signifying a satisfactory fit.
    
\begin{figure}[htbp]
    \centering
    \includegraphics[width=0.45\textwidth]{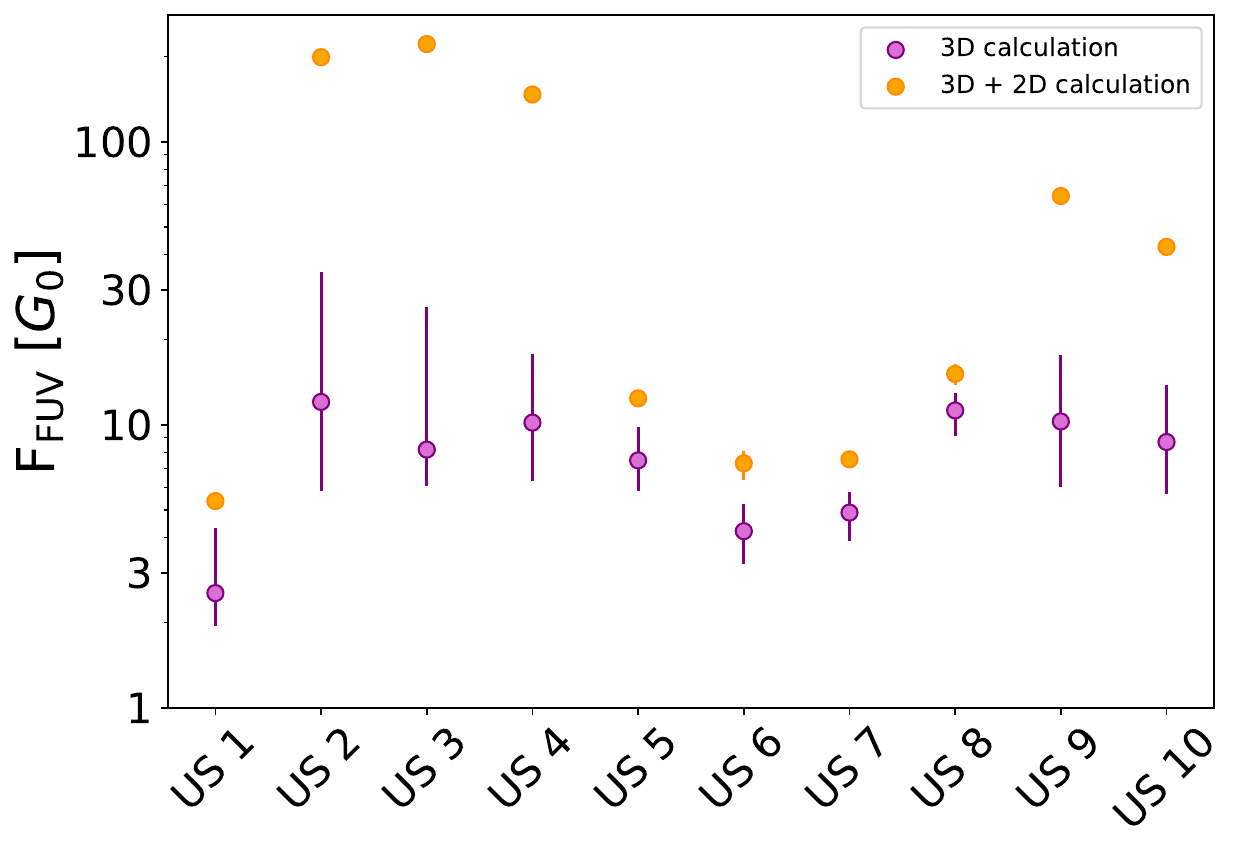}
    \caption{FUV fluxes and uncertainties for the ten AGE-PRO targets in Upper Sco, evaluated using Monte Carlo sampling of parallaxes (\textit{purple}, same as in Figure \ref{fig:fuv_flux}), and 2D projected separation between Upper Sco OB stars and the AGE-PRO disks (\textit{orange}).}
    \label{fig:fuv_flux_2d}
    \end{figure}

    It is important to highlight that in our utilization of the Monte Carlo method, we randomly selected values from a normal distribution, despite the fact that the distribution may not be perfectly Gaussian. 
    We can provide an upper limit of the FUV flux using the following method. We consider the OB-type stars members of Upper Sco \citep{Luhman_uppersco_census} and, for the closest pairs disk-massive star showing uncertainties on line-of-sight distances comparable or larger than their separation, we use the 2D projected separation (and the average distance to Upper Sco, 142 pc) to evaluate the flux as in equation \eqref{eq:fuv_flux}. For the other pairs we use the same Monte Carlo sampling method described in Section \ref{sec: FUV calculation}. In Figure \ref{fig:fuv_flux_2d} we compare the resulting FUV flux values obtained from applying the two calculation methods described above to the AGE-PRO targets in Upper Sco. The best estimate of the FUV flux is presumably between the two provided estimates. Adopting the flux upper limits in our simulations, instead of the median values, would result in more dispersed disks (less massive and extended). Consequently, higher initial disk masses would be required to explain the observations. However, as disks shrink slowly when the equilibrium between the mass lost in wind and the mass accreting onto the star is reached, we expect the final sizes not to be significantly smaller than what predicted by our models.  

\clearpage

\bibliography{sample631}{}
\bibliographystyle{aasjournal}



\end{document}